\documentclass{aa}

\usepackage{graphicx}
\usepackage{orcidlink}
\hypersetup{hidelinks}
\usepackage{txfonts}
\usepackage{subcaption}
\usepackage{lscape}
\usepackage{placeins}

\newcommand{\msolar}{M_{\rm \odot}}
\usepackage{multirow}
\usepackage{tabularx}
\usepackage{amssymb}
\usepackage{amsmath}
\newcommand{\ang}{\mathring{A}}

\makeatletter
\renewcommand*\aa@copyrightname{\copyright~The Authors}
\renewcommand*\aa@manuscriptname{manuscript no. \jobname\hspace{\stretch{1}}\aa@copyrightname~\the\year}
\makeatother

\begin{document}

   \title{X-ray measurements of the elemental abundances for the diffuse emission in the nuclear starburst galaxy NGC 3079}

   \author{Jiejia Liu\inst{1}\thanks{liujj21@mails.tsinghua.edu.cn}\,\orcidlink{0009-0003-8288-3985}
        \and Junjie Mao\inst{1}\thanks{jmao@mail.tsinghua.edu.cn}\,\orcidlink{0000-0001-7557-9713}
        \and Shuinai Zhang\inst{2,3}\,\orcidlink{0000-0002-8329-6606}
        \and Guan-Fu Liu\inst{1}\,\orcidlink{0000-0002-3464-4424}
        \and Rui Huang\inst{1,4}\,\orcidlink{0000-0001-7900-4204}
        \and Wei Cui\inst{1}\,\orcidlink{0000-0002-6324-5772}
        }

   \institute{Department of Astronomy, Tsinghua University, Beijing 100084, China
   \and Key Laboratory of Dark Matter and Space Astronomy, Purple Mountain Observatory, Chinese Academy of Sciences, Nanjing 210023, China
   \and School of Astronomy and Space Science, University of Science and Technology of China, Hefei 230026, China
   \and Department of Astronomy, University of Michigan, 311 West Hall, 1085 S. University Ave, Ann Arbor, MI 48109-1107, USA
   }

   \date{Received 27 March 2026 / Accepted 28 September 2026}

  \abstract
   {Outflows driven by feedback leave imprints on a galaxy's diffuse X-ray emission. Elemental abundances provide key diagnostics of feedback-driven chemical enrichment.}
   {We present a spectroscopic study of the diffuse X-ray emission in NGC~3079 using \textit{XMM-Newton} RGS and EPIC, focusing on the nuclear region and the extended galactic-scale superbubble (GSB).}
   {We analyzed the high-resolution RGS spectrum of the central $\sim4~\rm kpc$ region, encompassing the $\sim1~\rm kpc$ nuclear superbubble, and the EPIC spectrum of the GSB out to $5'~(\sim25~\rm kpc)$.}
   {The nuclear spectrum is best described by a multi-temperature thermal-plasma model plus a charge-exchange component, which contributes approximately $13\%$ of the total observed energy flux in the $0.2$--$2~\rm keV$ band. The best-fit O/Fe and Ne/Fe ratios are $0.47$ and $0.77$ in solar units. Comparing these ratios with the yields of initial-mass-function-weighted core-collapse supernovae gives an upper progenitor mass cutoff of $M_{\rm u}\simeq13$--$15~\msolar$ for pure core-collapse supernova enrichment. For the best-fit spectral model, including a Type Ia supernova contribution increases the inferred cutoff, but its $1\sigma$ upper limit remains below $25~\msolar$ for Type Ia supernova fractions up to $10\%$ within the adopted yield models. The characteristic ages of the nuclear superbubble and GSB are $0.92^{+0.39}_{-0.46}~\rm Myr$ and $29.6^{+1.6}_{-2.9}~\rm Myr$, respectively, suggesting episodic feedback in NGC~3079.}
   {}


    \titlerunning{Elemental abundances in NGC~3079}
   \maketitle
   \nolinenumbers

\section{Introduction}
Feedback processes play a crucial role in galaxy formation and evolution \citep{1998A&A...331L...1S, 1998AJ....115.2285M, Tumlinson_CGM_2017,2023ARA&A..61..131F}. The energy sources of feedback include stellar winds and supernova (SN) explosions (stellar feedback) as well as active galactic nuclei (AGNs; AGN feedback; \citealt{Silk_Mamon_2012}). These energetic outbursts inject mass, energy, and momentum into the interstellar medium (ISM) and the circumgalactic medium (CGM), leaving imprints on a galaxy's atmosphere. They can launch supersonic winds and produce X-ray superbubbles and cavities with temperatures of $\sim10^6~\rm K$, extending to tens of kiloparsecs above both sides of the galactic disk, as seen in simulations \citep{Pillepich_2021_TNG}. Observations also show that superbubbles are ubiquitous in galaxies with intense star formation (e.g., starburst galaxies) and AGN activity \citep{Strickland_2004_outflow,Li_Wang_2013_xraycoronae_discgal} and in relatively quiescent galaxies such as the Milky Way \citep{Snowden_1997_ROSAT_SXRB, Predehl_2020}.
Observationally, a characterization of the emission lines from the superbubble reveals their chemical abundance and the thermal and dynamical states, and it provides insights into the feedback processes.

NGC 3079 is an SB(s)c starburst galaxy \citep{Lawrence_1985_3079starburst, Veilleux_1994_3079starburst} with a nuclear star formation rate (SFR) of $\sim 2.6~\msolar~\rm yr^{-1}$ \citep{Yamagishi_2010_3079_starburst}. Outflows leave imprints at different spatial scales in NGC 3079. Within the nuclear region, a $\sim1~\rm kpc$ nuclear superbubble (NSB) is detected in radio, optical, and X-ray bands. On larger scales, a pair of galactic superbubbles (GSB) extends up to $\sim25~\rm kpc$ from the nucleus, with an X-ray luminosity of $6\times10^{39}~\rm erg~s^{-1}$ \citep{Pietsch_1998}. An additional far-ultraviolet component associated with the GSB extends farther out to $\sim60~\rm kpc$ \citep{Hodges-Kluck_2020}.
Abundance measurements reveal differences between different observations and theoretical predictions. \citet{Konami_2012, Li_2019ApJ_3079} measured the O/Fe and Ne/Fe ratios in the nuclear region using Suzaku and found consistency with yields from core-collapse supernovae (SNcc) weighted by the initial mass function (IMF), suggesting SNcc enrichment. In contrast, \citet{Hodges-Kluck_2020} analyzed the \textit{XMM-Newton} European Photon Imaging Camera (EPIC) data and reported a central O/Fe ratio higher by $\sim 1\sigma$ than the theoretical SNcc prediction. For the GSB, \citet{Konami_2012} found that the O/Fe and Ne/Fe ratios decreased below the SNcc yields at $\sim2'$, while \citet{Hodges-Kluck_2020} reported an O/Fe ratio consistent with the theoretical prediction within the uncertainties.

We present a detailed study of the elemental abundances of the nuclear region and the GSB in NGC 3079. We measured the abundances using data from the \textit{XMM-Newton} Reflection Grating Spectrometer \citep[RGS;][]{den_Herder_2001_xmm_rgs} and EPIC \citep{Turner2001_xmm_epic_mos,Struder_2001_xmm_epic_pn}.
We compare the observed abundance ratios in detail to the SNcc theoretical models.
Throughout this paper, we adopted a distance of NGC 3079 of $17.3~\rm Mpc$ \citep[][redshift $z\approx0.004$, $1'\sim5~\rm kpc$]{1992ApJS...80..479T} and a proto-solar abundance from \cite{Lodders_2009} unless stated otherwise.

\begin{table*}
    \footnotesize
    \centering
    \caption{X-ray observations of NGC~3079.}
    \begin{tabular*}{\textwidth}{@{\extracolsep{\fill}}clcccccc@{}}
    \hline
    \hline
    Telescope & ObsID & Start date & Duration (ks) & \multicolumn{4}{c}{GTI (ks)} \\
    \cline{5-8}
    & & & & MOS1 & MOS2 & PN & RGS \\
    \hline
    \multirow{12}{*}{\textit{XMM-Newton}} & 110930201 & 2001-04-13 & 25.3 &  8.3 & 8.1 & 4.8 & 15.7 \\
    & 147760101 & 2003-10-14 &  44.4 &  14.9 & 20.5 & 7.3 & 34.7 \\
    & 802710101 & 2017-11-01 &  22.8 &  19.4 & 19.5 & 17.7 & 16.4 \\
    & 802710201 & 2017-11-03 &  22.4 &  9.5 & 10.1 & 7.6 & 12.6 \\
    & 802710301 & 2017-11-05 &  22.4 & 16.6 & 18.4 & 9.6 & 16.6 \\
    & 802710401 & 2017-11-09 &  22.4 & 5.6 & 6.2 & 3.9  & 17.7 \\
    & 802710501 & 2017-11-15 &  22.4 & 15.8 & 16.8 & 11.9 & 14.7 \\
    & 802710601 & 2017-11-23 &  22.4 & 14.5 & 15.1 & 0.2 & 14.7 \\
    & 802710701 & 2017-11-27 &  22.4 &  16.8 & 19 & 13.6 & 17.5 \\
    & 802710801 & 2018-04-17 &  26 &  24.6 & 24.6 & 22.3 & 18.6 \\
    & 802710901 & 2018-04-21 &  22.4 & 21 & 19.2 & 21.2 & 16.3 \\
    & 802711001 & 2018-04-23 &  25.3 & 24 & 23.8 & 22.1 & 21.1 \\
    \hline
    \multirow{3}{*}{\textit{Chandra}} & 2038 &  2001-05-07 & 26.6 & \multicolumn{4}{c}{23.5} \\
    & 19307 & 2018-01-30 & 53.2 & \multicolumn{4}{c}{50.1} \\
    & 20937 & 2018-02-01 & 44.5 & \multicolumn{4}{c}{42.2} \\
    \hline
    \end{tabular*}
    \tablefoot{The telescope names, ObsIDs, observation start dates, durations, and total clean exposure times are shown. For \textit{XMM-Newton}, the GTIs for MOS1, MOS2, PN, and RGS are listed separately.}
    \label{tab:obs}
\end{table*}

\section{Observations and data reduction}
\label{sec:X-ray observations}
\subsection{X-ray observations}
We used \textit{XMM-Newton} and \textit{Chandra} archival observations in this paper. They are summarized in Table~\ref{tab:obs}. The primary observations we used are the observations of NGC~3079 under project 080271 (principal investigator: Hodges-Kluck). Two additional observations, with observation identifiers (ObsIDs) 0110930201 and 0147760101, are also included.
The data were processed with the Science Analysis System (SAS; version 20.0.0). We used \texttt{cifbuild} to generate the current calibration files (CCF) and \texttt{odfingest} to extend the observation data file (ODF) summary file. The tasks \texttt{epchain} and \texttt{emchain} were used to generate the event-list product for the EPIC metal-oxide-semiconductor (MOS) and pn cameras. The task \texttt{emanom} was used to examine the MOS charge-coupled devices (CCDs) for anomalous states where the background at $E < 1~\rm keV$ is strongly enhanced \citep{Kuntz_Snowden_2008}. The soft proton (SP) flare filtering was accomplished by the tasks \texttt{mos-filter} and \texttt{pn-filter}, respectively. We note that observations with ObsID 0110930201 and 0147760101 are severely contaminated by soft protons. The total filtered exposure time is 191.0 ks, 201.3 ks, and 142.2 ks for MOS1, MOS2, and pn, respectively.

The \textit{Chandra} data were processed with the Chandra Interactive Analysis of Observations package (CIAO; version 4.14; \citealt{Fruscione2006_ciao}) and the calibration database (CALDB) version 4.9.6. The observations were reprocessed using the tool \texttt{chandra\_repro}. Background flares beyond $3\sigma$ were filtered out using \texttt{deflare} with the script \texttt{lc\_clean}. The accumulated good time intervals (GTIs) for the three observations are 115.8 ks.

\subsection{RGS spectra for the nuclear region}
\begin{figure*}
    \centering
    \includegraphics[width=0.95\linewidth,trim={0.3cm 0cm 0cm 0cm}]{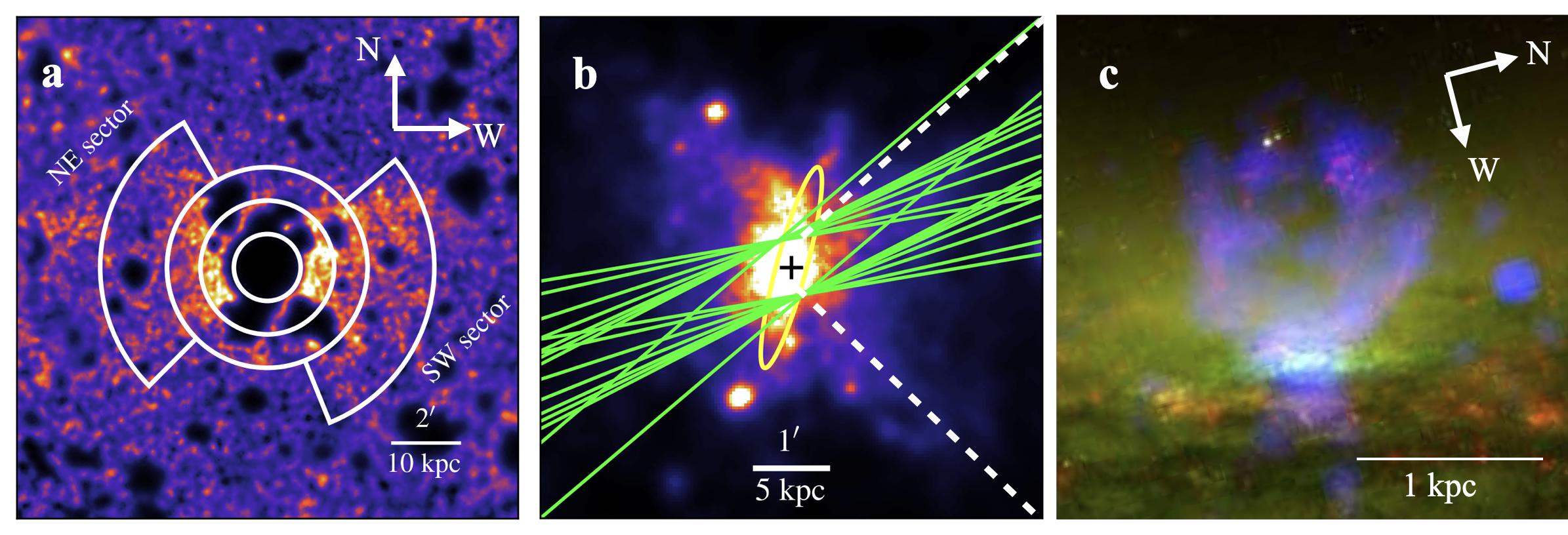}
    \caption{\textit{Panels a and b}: EPIC-MOS and RGS source extraction regions overlaid on the mosaicked, background-subtracted, exposure-corrected \textit{XMM-Newton} EPIC images. In panel a, the adaptively smoothed (50-count threshold) point-source-masked image in the $0.3$--$2.0~\rm keV$ band is shown, together with the annulus and sector regions used for the GSB spectra. In panel b, the adaptively smoothed (100-count threshold) image in the $0.45$--$1.25~\rm keV$ band is shown. \textit{Panel c}: Zoomed-in false-color image of the nuclear region, where red, green, and blue correspond to the Hubble Space Telescope I band, N {\sc ii}+H$\alpha$, and the \textit{Chandra} X-ray image, respectively.}
    \label{fig:rgs_region}
\end{figure*}

We used all observations listed in Table~\ref{tab:obs} to extract the high-resolution X-ray spectra of the NSB using \textit{XMM-Newton} RGS in the $8$--$26~\rm\ang$ range ($\sim0.5$--$1.6~\rm keV$), following the procedure described by \citet{JMao_2023_RGS_instru}. The event files were generated using the \texttt{rgsproc} pipeline and by specifying the source coordinates $(\rm ra,dec)=(150.49117, 55.679848)$. The parameter \texttt{xpsfincl} was set to 90 to include 90\% of the point spread function (PSF) of the source region, corresponding to a cross-dispersion width of $\sim0.8'$ (Fig.~\ref{fig:rgs_region}b). The background light curve was extracted in the wavelength range $\lambda \sim5$--$7~\rm \ang$ on CCD 9 with a time bin of $100~\rm s$. We filtered background flares using the \texttt{deflare} task from the CIAO package, applying a $3\sigma$ threshold. For ObsIDs 147760101 and 802710401, which were strongly affected by background flares, a more conservative $2\sigma$ threshold was adopted. After filtering, the total clean exposure time of all observations was 216.6 ks. The data were reprocessed with the filtered GTIs using \texttt{rgsproc}. The RGS1 and RGS2 spectra from all observations were combined using \texttt{rgscombine} with the modeled background subtracted, as shown in Fig.~\ref{fig:rgs_spec_bestfit}. The spectrum shows prominent Ne {\sc x} Ly$\alpha$, O {\sc viii} Ly$\alpha$, and Fe-L shell emission lines.

NGC 3079 hosts a Compton-thick AGN with a hydrogen column density of $N_{\rm H}\sim10^{24}~\rm cm^{-2}$ \citep{Iyomoto_2001_ApJL, LaCaria_2019}, making its contribution negligible in the RGS band ($\lambda=8$--$26~\rm \ang$). Nevertheless, to quantify its contribution, we extracted the \textit{Chandra} spectrum with \texttt{specextract} from a $1''$ circular region centered at $(\rm ra,dec)=(150.49083,55.67987)$ and combined all observations listed in Table~\ref{tab:obs} using \texttt{combine\_spectra}.

\subsection{EPIC spectra for GSB}
We performed a spatially resolved spectral analysis for the GSB. The source extraction regions are shown in Fig.~\ref{fig:rgs_region}a. The regions were defined as annuli centered at $(\rm RA,Dec) = (150.49116,55.67985)$, with inner and outer radii of $1'$-$2'$ and $2'$-$3'$, respectively. For the outermost regions, sector-shaped extraction areas were used to reduce the contamination from the Galactic foreground. The northeast (NE) and southwest (SW) sectors were defined between $3'$ and $5'$ with an angular span of $105^\circ$ and $108^\circ$, respectively, and were combined to improve the spectral statistics. For extended sources, spatial variations in the ancillary response file and redistribution matrix file cannot be neglected.
To account for the sky X-ray background, we extracted and fit a spectrum from an annulus between $6'$ and $10'$, where the GSB surface brightness has declined to the background level \citep{Hodges-Kluck_2020}, then rescaled the best-fit model as the background components in the following spectral fitting (see details in Sect.~\ref{sec:spec_analysis_gsb}).

We adopted the thirteenth data release of the fourth \textit{XMM-Newton} Serendipitous Source Catalogue (4XMM-DR13)\footnote{\url{http://xmmssc.irap.omp.eu/Catalogue/4XMM-DR13/4XMM_DR13.html}} as the source list and generated the source exclusion regions using the task \texttt{region}. The exclusion regions were defined using the contour method, in which elliptical regions are determined by a contour level equal to 0.4 times the local background (\texttt{bkgfraction}=0.4). Since the PSF broadens with increasing off-axis angle,
a more conservative setting of \texttt{bkgfraction}=0.1 was applied to the NE and SW sectors and to the region used for foreground modeling. We also manually added masks after visual inspection.

We extracted spectra using only MOS1 and MOS2. The pn unexposed area is relatively small, making the particle-background prediction more uncertain \citep{Bulbul_2020_xmmpnbkg}, and pn data have recently been reported to be contaminated by soft protons \citep{Bulbul_2020_xmmpnbkg,Marelli_2021ApJ_xmmpnbkg}. We excluded ObsIDs 0110930201, 0147760101, and 0802710601 because they are severely contaminated by particle background. For the NE+SW sector, we used MOS2 alone because the quality of MOS1 was insufficient. We extracted MOS spectra from the regions in Fig.~\ref{fig:rgs_region}a using \texttt{mos-spectra}. The quiescent particle background \citep[QPB;][]{Kuntz_Snowden_2008} was generated with \texttt{mos\_back} and was modeled as an additive component (i.e., not subtracted). Spectra from individual observations were then combined for MOS1 and MOS2 separately using \texttt{epicspeccombine}.

For each observation, we used \texttt{mos-spectra} to generate point-source-masked images and exposure maps in the $0.3-2~\rm keV$ band, and \texttt{mos\_back} to generate the corresponding QPB images. The mosaicked, QPB-subtracted, and exposure-corrected image was produced with \texttt{merge\_obs\_xmm} and was subsequently smoothed using \texttt{merge\_adapt} with a threshold count of 50. The final image is shown in Fig.~\ref{fig:rgs_region}a.

\section{Spectral analysis}
\label{sec:spec_analysis}
\begin{figure*}
    \centering
    \includegraphics[width=0.9\linewidth,trim={0.2cm 0.04cm 0.8cm 0.06cm},clip]{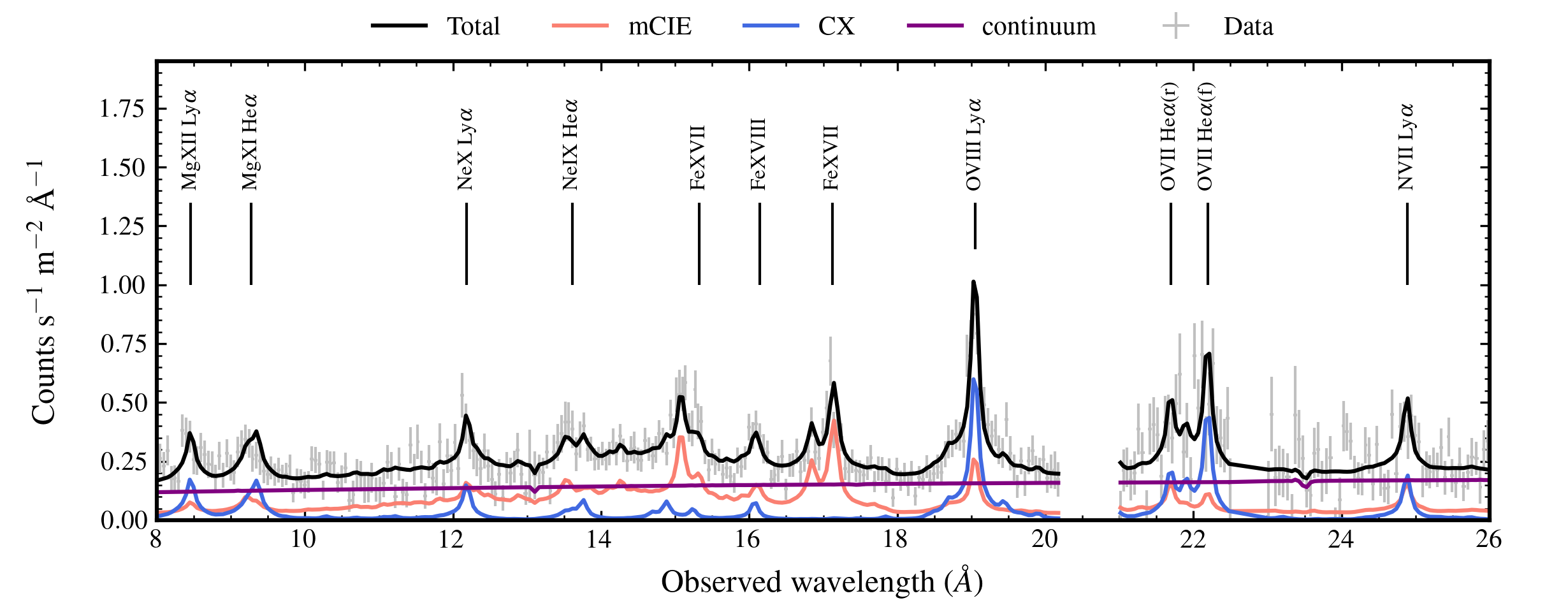}
    \caption{Best-fit model to the X-ray RGS spectrum of the nuclear region. The RGS1 and RGS2 combined data are plotted using silver crosses, and prominent lines are labeled. The black curve shows the best-fit model. The mCIE, CX, and continuum components are shown by the salmon, blue, and purple curves, respectively. The AGN component is not plotted since it is negligible in this band.}
    \label{fig:rgs_spec_bestfit}
\end{figure*}
We performed the spectral analysis using SPEX \citep[v3.08.01;][]{Kaastra_1996_SPEX,Kaastra_zenodo_SPEX_v3.08.01}, where the most recent collision data for H- and He-like lines \citep{JMAO_2022ApJS} are included. C-statistic \citep{Kaastra_2017A&A_Cstat} was used for the X-ray spectral fitting. We fixed the redshift of NGC~3079 at $z=0.004$ \citep{1992ApJS...80..479T} and adopted cosmological parameters $H_0=70~{\rm km~s^{-1}~Mpc^{-1}}$, $\Omega_M=0.3$, and $\Omega_\Lambda=0.7$. We included Galactic foreground absorption and fixed the neutral hydrogen column density at $N_{\rm H}=8.78\times10^{19}~\rm cm^{-2}$ \citep{HI4PI_2016} in all subsequent spectral fittings.

\subsection{Nuclear region}
\label{sec:spec_analysis_nsb}
We first quantified the AGN contribution by fitting the combined \textit{Chandra} spectrum from a $1''$ region centered on the supermassive black hole (SMBH) with \texttt{hot*red*(hot*powerlaw+refl)}. Here, \texttt{refl} modeled the reflected continuum and associated line emission (dominated by Fe K at 6.4 keV; \citealt{Magdziarz_1995MNRAS_refl, Zycki_1999MNRAS_refl}), and the two \texttt{hot} components accounted for Galactic and local absorption (the latter was fixed at $N_{\rm H}=2.5\times10^{24}~\rm cm^{-2}$; \citealt{LaCaria_2019}). We obtained $\Gamma=1.38^{+0.15}_{-0.13}$ and an AGN flux of $\approx1.5\times10^{-15}~\rm erg~cm^{-2}~s^{-1}$ in $7.5$--$26~\rm\ang$, that is, $\sim100$ times below the nuclear-region flux, and thus, negligible.

We adopted the best-fit AGN spectral model as a fixed component (\texttt{AGN}) in the subsequent analysis. We fit the RGS spectrum with a multi-temperature collisional ionization equilibrium (CIE) component, denoted mCIE, together with charge-exchange (CX) and continuum components. Model M1 is \texttt{mCIE+CX+pow+AGN}, as described below.
\begin{itemize}
    \item \texttt{mCIE} is a multi-temperature CIE component \citep{Kaastra_1996_SPEX} with a log-normal temperature distribution\footnote{\url{https://spex-xray.github.io/spex-help/models/cie.html}} (central temperature $kT_0$ and width $\sigma_T$). For M1, we constrained the temperature-distribution width to $\sigma_T\leq1$. The O, Ne, Mg, and Fe abundances were free parameters.
    \item \texttt{CX} is a CX component \citep{Gu_2016A&A_cx} motivated by the prominent O \textsc{vii} He$\alpha$ forbidden line at a rest wavelength of $22.10~\rm\ang$ (Fig.~\ref{fig:rgs_spec_bestfit}). We coupled $kT_{\rm CX}$, $\sigma_{\rm CX}$, and the abundances to \texttt{mCIE}, assuming that CX and CIE arise from the same hot plasma. We assumed that the hot-cold interaction is dominated by outflow velocity (\texttt{mode=2}) and allowed the collision velocity to vary, obtaining $v_{\rm CX}=264^{+141}_{-96}~\rm km~s^{-1}$. We used neutral H as the reference target (\texttt{ref=1}), a target density of approximately $1~\rm cm^{-3}$, the optically thin option (\texttt{op=1}), and the default $l$-distribution weighting (\texttt{wt=1}).
    \item \texttt{pow} is a power-law component accounting for the continuum from unresolved stellar sources and nonthermal emission from the southwest NSB \citep{Li_2019ApJ_3079}.
\end{itemize}

All components were redshifted and absorbed by Galactic foreground gas with fixed $N_{\rm H}$. To account for spatial broadening in the RGS, the spectral model was convolved by \texttt{lpro} with a profile generated by \texttt{rgsvprof} \citep{Tamura_2004_lpro}. The profile was derived from the MOS detector image in the $0.45$--$1.25~\rm keV$ band using the full $-5'$ to $+5'$ dispersion range.

The fitted parameters with $1\sigma$ errors are shown in Table~\ref{tab:fitpar}. The \texttt{mCIE} and \texttt{CX} contribute approximately 26\% and 13\%, respectively, to the observed $0.2$--$2~\rm keV$ energy flux. The Mg abundance is poorly constrained. We further discuss the abundance ratios and their systematics in Sect.~\ref{sec:discussion-model}.

\begin{table*}
    \centering
    \caption{Spectral fitting results.}
    \fontsize{6.8}{8.2}\selectfont
    \setlength{\tabcolsep}{2.2pt}
    \renewcommand{\arraystretch}{1.12}
    \begin{tabular*}{\textwidth}{@{\extracolsep{\fill}}lcccccccc@{}}
    \hline
    \hline
    Parameters & \multicolumn{5}{c}{Nuclear region (RGS)} & \multicolumn{3}{c}{GSB (EPIC-MOS)} \\
    \cline{2-6}\cline{7-9}
    & M1 & M2 & M3 & M4 & M5 & $1'-2'$ & $2'-3'$ & NE+SW \\
    Model & mCIE+CX & NEIJ+mCIE & mCIE & 2CIE+CX & CIE+CX & CIE & CIE & CIE \\
    \hline
    $kT_{\rm CIE}$ (keV) & $0.47^{+1.41}_{-0.24}$ & $0.38^{+0.06}_{-0.05}$ & $0.46^{+0.54}_{-0.20}$ & $0.29^{+0.05}_{-0.04}$ & $0.49^{+0.03}_{-0.04}$ & $0.38^{+0.01}_{-0.01}$ & $0.35^{+0.07}_{-0.01}$ & $0.30^{+0.07}_{-0.03}$ \\
    $\sigma_T$ & $0.76^{+0.24}_{-0.27}$ & $0.42^{+0.23}_{-0.12}$ & $1^{+0}_{-1}$ & -- & -- & 0 (fixed) & 0 (fixed) & 0 (fixed) \\
    $kT_{\rm NEIJ}/kT_{\rm high}$ (keV) & -- & $0.10^{+0.03}_{-0.02}$ & -- & $0.74^{+0.10}_{-0.11}$ & -- & -- & -- & -- \\
    $U_{\rm NEIJ}$ ($10^{20}~\rm s\,m^{-3}$) & -- & $(5.58^{+4.15}_{-2.48})\times10^{-6}$ & -- & -- & -- & -- & -- & -- \\
    $kT_{\rm CX}$ (keV) & coupled & -- & -- & $0.21^{+0.01}_{-0.01}$ & $0.22^{+0.01}_{-0.01}$ & -- & -- & -- \\
    $v_{\rm CX}$ ($\rm km\,s^{-1}$) & $264^{+141}_{-96}$ & -- & -- & $120^{+70}_{-120}$ & $134^{+480}_{-134}$ & -- & -- & -- \\
    O ($\rm O_\odot$) & $0.25^{+0.06}_{-0.03}$ & $0.43^{+0.12}_{-0.08}$ & $0.63^{+0.23}_{-0.14}$ & $0.26^{+0.04}_{-0.03}$ & $0.25^{+0.04}_{-0.03}$ & $0.69^{+0.15}_{-0.13}$ & $0.72^{+0.24}_{-0.18}$ & $0.37^{+0.11}_{-0.09}$ \\
    Ne ($\rm Ne_\odot$) & $0.41^{+0.14}_{-0.11}$ & $0.52^{+0.19}_{-0.14}$ & $0.79^{+0.35}_{-0.23}$ & $0.36^{+0.10}_{-0.08}$ & $0.44^{+0.36}_{-0.08}$ & $0.61^{+0.13}_{-0.10}$ & $0.62^{+0.22}_{-0.17}$ & $0.22^{+0.16}_{-0.22}$ \\
    Mg ($\rm Mg_\odot$) & $1.49^{+0.55}_{-0.43}$ & $1.98^{+0.89}_{-0.62}$ & $3.86^{+1.41}_{-0.91}$ & $3.13^{+1.45}_{-1.08}$ & $1.61^{+1.53}_{-0.44}$ & 1 (fixed) & 1 (fixed) & 1 (fixed) \\
    Fe ($\rm Fe_\odot$) & $0.53^{+0.19}_{-0.13}$ & $0.28^{+0.08}_{-0.06}$ & $0.40^{+0.13}_{-0.08}$ & $0.33^{+0.13}_{-0.09}$ & $0.24^{+0.19}_{-0.05}$ & $0.20^{+0.04}_{-0.03}$ & $0.14^{+0.05}_{-0.04}$ & $0.09^{+0.06}_{-0.04}$ \\
    $\mathrm{Norm}_{\rm CIE}$ ($10^{64}~\rm m^{-3}$) & $(9.8^{+12}_{-4.5})\times10^{4}$ & $(1.4^{+0.32}_{-0.25})\times10^{5}$ & $(1.7^{+0.45}_{-0.43})\times10^{5}$ & $(5.6^{+1.5}_{-1.3})\times10^{4}$ & $(6.9^{+1.9}_{-1.7})\times10^{4}$ & -- & -- & -- \\
    $\mathrm{Norm}_{\rm high}$ ($10^{64}~\rm m^{-3}$) & -- & -- & -- & $(3.2^{+1.2}_{-0.74})\times10^{4}$ & -- & -- & -- & -- \\
    $\mathrm{Norm}_{\rm NEIJ}$ ($10^{64}~\rm m^{-3}$) & -- & $(2.2^{+7.2}_{-1.7})\times10^{6}$ & -- & -- & -- & -- & -- & -- \\
    $\mathrm{Norm}_{\rm CX}$ ($10^{64}~\rm m^{-3}$) & $(6.1^{+6.9}_{-3.1})\times10^{5}$ & -- & -- & $(3.6^{+7.1}_{-2.2})\times10^{5}$ & $(4.6^{+8.9}_{-3.7})\times10^{5}$ & -- & -- & -- \\
    \hline
    \multicolumn{9}{c}{$0.2$--$2~\rm keV$ luminosity ($\rm erg\,s^{-1}$)} \\
    \hline
    $L_{X,\rm CIE}$ & $1.2\times10^{40}$ & $1.7\times10^{40}$ & $2.0\times10^{40}$ & $4.9\times10^{39}$ & $8.4\times10^{39}$ & $1.9\times10^{39}$ & $2.0\times10^{39}$ & $1.4\times10^{39}$ \\
    $L_{X,\rm NEIJ/high}$ & -- & $1.1\times10^{41}$ & -- & $6.3\times10^{39}$ & -- & -- & -- & -- \\
    $L_{X,\rm CX}$ & $5.1\times10^{39}$ & -- & -- & $1.9\times10^{39}$ & $2.6\times10^{39}$ & -- & -- & -- \\
    $L_{X,\rm pow}$ & $2.8\times10^{40}$ & $2.1\times10^{40}$ & $2.4\times10^{40}$ & $2.7\times10^{40}$ & $2.8\times10^{40}$ & -- & -- & -- \\
    \hline
    \multicolumn{9}{c}{Fit statistics} \\
    \hline
    $C_{\rm exp}$ & 337.2 & 337.2 & 337.2 & 337.3 & 337.3 & 521 & 529 & 252 \\
    $C_{\rm stat}$ & 410.6 & 409.2 & 429.1 & 423.9 & 430.2 & 561 & 536 & 287 \\
    $\Delta C$ & 26 & 26 & 26 & 26 & 26 & 33 & 33 & 26 \\
    $\rm AIC_{\rm c}$ & 433.4 & 434.2 & 447.6 & 451.0 & 453.0 & -- & -- & -- \\
    $\Delta\rm AIC_{\rm c}$ & 0.0 & 0.8 & 14.3 & 17.7 & 19.6 & -- & -- & -- \\
    \hline
    \end{tabular*}
    \tablefoot{Fitted parameters and $1\sigma$ errors for the nuclear region RGS spectrum and the GSB EPIC-MOS spectra in the $1'-2'$, $2'-3'$ annuli and the NE+SW sectors. The five nuclear region models are M1 \texttt{mCIE+CX+pow}, M2 \texttt{NEIJ+mCIE+pow}, M3 \texttt{mCIE+pow}, M4 \texttt{2CIE+CX+pow}, and M5 \texttt{CIE+CX+pow}. All nuclear region models include Galactic absorption, redshift, RGS spatial broadening, and the fixed AGN component. For M4, $kT_{\rm CIE}$ and $kT_{\rm high}$ denote the low- and high-temperature CIE components, respectively. The GSB spectra were fitted with a single CIE component. Abundances are relative to solar values. SPEX normalizations are given in units of $10^{64}~\rm m^{-3}$: the CIE and NEIJ normalizations are emission measures $n_{\rm H}n_{\rm e}V$, while the CX normalization is $n_{\rm H}n_{\rm nh}V$. The $0.2$--$2~\rm keV$ luminosities are listed for the fitted thermal, NEIJ, CX, and soft power-law components; the fixed AGN and reflection components are omitted. The C-statistics, corrected AIC $\rm AIC_{\rm c}$ and $\Delta \rm AIC_c$ are also shown.}
    \label{tab:fitpar}
\end{table*}

We also tested four alternative models: M2 (a nonequilibrium ionization (NEI) component, \texttt{NEIJ}, plus \texttt{mCIE} and a continuum; \texttt{NEIJ+mCIE+pow}), M3 (a multi-temperature \texttt{CIE} component plus a continuum without CX; \texttt{mCIE+pow}), M4 (a two-temperature \texttt{CIE} model plus CX and a continuum; \texttt{2CIE+CX+pow}), and M5 (a single-temperature \texttt{CIE} model plus CX and a continuum; \texttt{CIE+CX+pow}). All models included redshift, Galactic absorption, RGS line broadening, and the fixed AGN component. For M4 and M5, we allowed $v_{\rm CX}$ to vary and obtained $120^{+70}_{-120}$ and $134^{+480}_{-134}~\rm km~s^{-1}$, respectively. The confidence intervals reach zero and have large upper uncertainties, indicating that our data constrain $v_{\rm CX}$ only weakly in these models. The results are listed in Table~\ref{tab:fitpar}.

We compared models M1--M5 using the Akaike information criterion (AIC) corrected for finite sample size \citep{AIC_1974,corrected_AIC_Sugiura_1978},
\begin{equation}
    {\rm AIC_c} = {\rm AIC} + \frac{2k(k+1)}{N-k-1}
    \label{eq:corrected_aic}.
\end{equation}
Here, $N$ is the number of data points, and $k$ is the number of free parameters in the model. We report ${\rm AIC_c}$ and $\Delta\mathrm{AIC}_\mathrm{c}=\mathrm{AIC}_\mathrm{c}-\mathrm{AIC}_{\mathrm{c,min}}$ in Table~\ref{tab:fitpar}. The preferred model is the one with the minimum ${\rm AIC_{c,min}}$. Models with $\Delta {\rm AIC_c} > 2$ are considered to be considerably less supported by the data \citep{burnham2002model}. M1 gives the minimum $\mathrm{AIC}_\mathrm{c}$, while M2 is statistically comparable ($\Delta\mathrm{AIC}_\mathrm{c}=0.8$); M3--M5 are considerably less supported. However, a literal single-ionization-age interpretation of the best-fit M2 solution implies an implausibly high mass-processing rate and heating power. We therefore adopted M1 as the fiducial model for the physical and chemical enrichment interpretation and discuss the physical consistency of M2 in Sect.~\ref{sec:discussion-driver}.

Although the Compton-thick AGN contributes negligibly to the continuum in the RGS band along our line of sight, leaked AGN photons might still photoionize the ambient gas and enhance the O~{\sc vii} (f) line. Since NGC~3079 is classified as both a Seyfert~2 galaxy and a low-ionization nuclear emission-line region \citep{Ford_1986_Seyfert2, Heckman_1980_LINER} and its AGN spectral energy distribution is poorly constrained, we did not include photoionization components in the spectral modeling.

\subsection{Galactic-scale superbubble}
\label{sec:spec_analysis_gsb}
We modeled the sky foreground and background using the $6'$--$10'$ blank-sky spectrum (Appendix~\ref{appendix:xray_background}). We rescaled the best-fit model by the area ratio of the source and background regions and included it as a fixed component in the subsequent spectral fitting of the GSB. The spectra extracted from the annular and sector regions (Fig.~\ref{fig:rgs_region}a) were modeled with a single \texttt{CIE} component (with thawed O, Ne, and Fe abundances), two delta functions for the instrumental Al K$\alpha$ and Si K$\alpha$ lines, and a broken power law for residual SP contamination. We did not adopt a multi-temperature \texttt{CIE} model for the GSB because the photon statistics are limited.

The fitted spectra for each region are shown in Fig.~\ref{fig:fit_epic}, and the best-fit parameters with the uncertainties are listed in Table~\ref{tab:fitpar}. The temperature of the hot plasma is consistent within the uncertainties out to $5'$, corresponding to $\sim25~\rm kpc$. We also fitted the $1'$--$2'$ annulus, which borders the disk and may contain hot--cold gas interfaces, with a \texttt{CIE+CX} model. The CX normalization converges to zero, which means that no CX component is required by the data, although it cannot be excluded at the CCD resolution (Sect.~\ref{sec:discussion-model}). For the NE+SW sectors, no cold gas has been detected out to $\sim25~\rm kpc$, and our spectra require no additional CX component. We use the GSB fits for the temperature and emission measure in Sect.~\ref{sec:discussion-driver}.

\section{Discussion}
\label{sec:discussion}
\subsection{CX contamination and multi-temperature plasma}
\label{sec:discussion-model}
\begin{figure}[t]
    \centering
    \includegraphics[width=\linewidth]{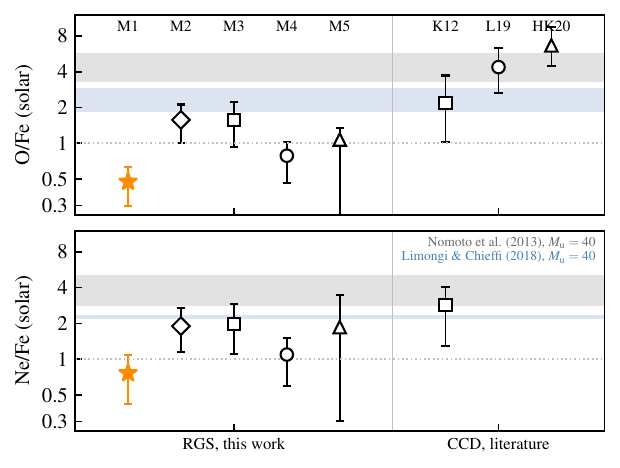}
    \caption{O/Fe (\textit{panel a}) and Ne/Fe (\textit{panel b}) in the nuclear region of NGC~3079 from the RGS models M1--M5 of this work (Table~\ref{tab:fitpar}) and from CCD spectra of the central region reported by \citealt[K12]{Konami_2012}, \citealt[L19]{Li_2019ApJ_3079}, \citealt[HK20]{Hodges-Kluck_2020} (O/Fe only for L19 and HK20). All values are rescaled to the proto-solar ratios of \citet{Lodders_2009}. The gray and blue bands show the IMF-weighted SNcc predictions for $M_{\rm u}=40~\msolar$ from \citet{NKT13_2013_SNcc} and \citet{LC18} over the initial metallicities of the yield tables (Sect.~\ref{sec:abund:prog}). The dotted line marks the solar ratio.}
    \label{fig:nuclear_ratio}
\end{figure}

Figure~\ref{fig:nuclear_ratio} compares the O/Fe and Ne/Fe ratios of our nuclear models with the CCD-based values reported for the central region and with the IMF-weighted SNcc predictions of \citet{NKT13_2013_SNcc} and \citet{LC18} for a conventional cutoff of $M_{\rm u}=40~\msolar$. The M1 ratios are several times lower than the CCD-based values. Two effects account for the difference. The first effect is CX. In M1, the \texttt{mCIE} and \texttt{CX} components share the same abundances, and the CX component contributes $13\%$ to the observed $0.2$--$2~\rm keV$ energy flux. CX contributes approximately $67\%$, $69\%$, and $52\%$ to the combined thermal and CX emission around the O~{\sc vii} triplet, O~{\sc viii} Ly$\alpha$, and Ne~{\sc x} Ly$\alpha$, respectively, but only $2$--$6\%$ around the Fe~{\sc xvii} lines. In this model, a large part of the O and Ne line emission is produced by CX, which lowers the fitted ratios: without CX (M3), they are $1.6$ and $2.0$, and in M1, they are $0.47$ and $0.77$. The CX component is motivated by the prominent O~{\sc vii} forbidden emission at a rest wavelength of $22.10~\rm\ang$ (Fig.~\ref{fig:rgs_spec_bestfit}) and supported by the improved fit relative to M3 ($\Delta{\rm AIC_c}=14$). The observed H~{\sc i} outflow \citep{Shafi_2015MNRAS_3079HI} and dust entrained in the wind \citep{Veilleux_2021_3079_IR} provide a plausible source of the cold gas required for CX. The statistically comparable M2 gives ${\rm O/Fe}\simeq1.6$, but its single-age NEI interpretation is physically implausible (Sect.~\ref{sec:discussion-driver}). The second effect is the temperature structure. In our simulations, a single-temperature fit to a multi-temperature plasma biases the $\alpha$/Fe ratios upward (Appendix~\ref{sec:bias}), and with CX included, the single-temperature model M5 gives ${\rm O/Fe}\simeq1.1$, which is twice the M1 value. The single-temperature \texttt{CIE+CX} model (M5) is worse than M1 by $\Delta{\rm AIC_c}=19.6$. We also tested replacing the CX component of M1 by a cool \texttt{CIE} component. This gave ${\rm O/Fe}=1.7$ but was worse than M1 by $\Delta{\rm AIC_c}=8$, with the cool component driven to the lower bound of its temperature range.

The GSB spectra cannot constrain either effect with the current observations. At the CCD resolution, the O~{\sc vii} triplet is unresolved, so a CX component can be neither required nor excluded, and the limited statistics do not support a multi-temperature model, whereas a single-temperature assumption is physically too strong for a multiphase outflow. For the GSB, the \texttt{CIE+CX} fit of the $1'$--$2'$ annulus does not require CX, and no cold gas has been detected at the outer extent of the NE+SW sectors (Sect.~\ref{sec:spec_analysis_gsb}). The current CCD spectra cannot exclude a CX contribution. We therefore do not discuss the CCD-based abundance patterns further.

The M1 ratios also lie well below the $M_{\rm u}=40~\msolar$ prediction, by a factor of $4$--$12$ for \citet{NKT13_2013_SNcc} and $3$--$6$ for \citet{LC18}. The difference between the observation and the theoretical prediction might arise from (1) uncertainties and simplifying assumptions in the SNcc yield models, (2) contributions from time-delayed Type Ia supernova (SNIa) enrichment, and (3) biases in the inferred abundance ratios due to CX contamination. For (3), we note that model M3, which does not include a CX component, still lies a factor of $2$--$4$ below the \citet{NKT13_2013_SNcc} prediction and at the lower edge of the \citet{LC18} range. We discuss (1) and (2) in the next section, where we examine which progenitor mass cutoff and SNIa fraction reproduce the M1 pattern, and we quantify (3) in Appendix~\ref{sec:discuss_cx} by repeating the comparison for the alternative CX prescriptions M3 to M5.

\subsection{Progenitor mass upper limit and SNIa fraction}
\label{sec:abund:prog}
\begin{figure*}
    \centering
    \includegraphics[width=0.90\linewidth]{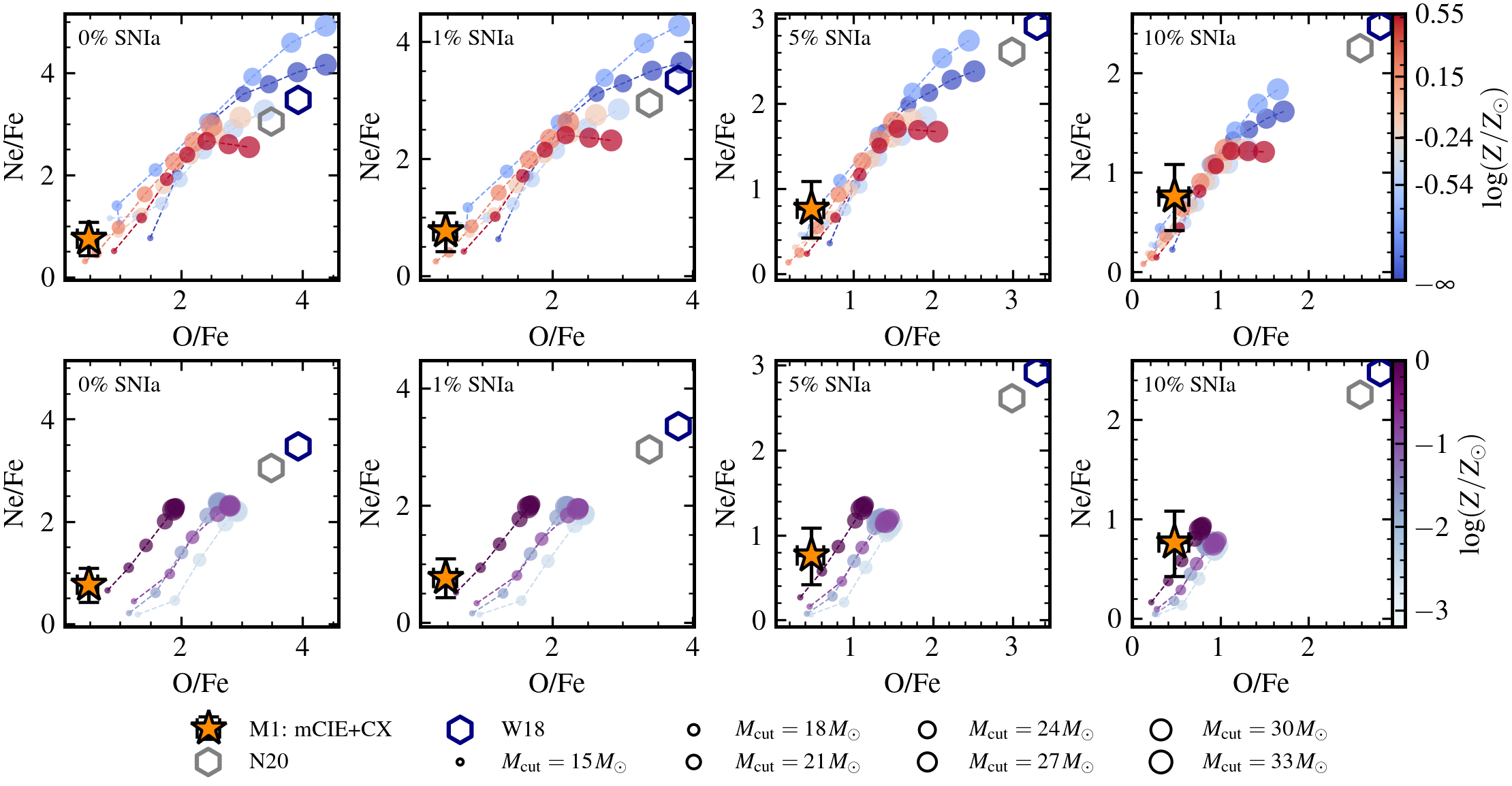}
    \caption{Comparison of the O/Fe and Ne/Fe ratios (in solar units) to theoretical predictions with varied initial metallicities, progenitor mass upper limits, and SNIa fractions. The fiducial M1 (mCIE+CX) measurement is shown as an orange star. The upper and lower panels adopt the SNcc yield tables from \citet{NKT13_2013_SNcc} and \citet{LC18}, respectively. The interpolation and extrapolation method of \citep{Gjergo_2023ApJS..264...44G} was used to compute IMF-weighted SNcc yields.
    The panels from left to right show the model predictions with 0\%, 1\%, 5\%, and 10\% SNIa contributions. The circles of different colors represent models with different initial metallicities, and the circle sizes correspond to progenitor mass cutoffs ranging from $15$--$33~\msolar$. The colored lines are drawn to guide the eye. In each panel, the SNcc model predictions from \citet{S16}, assuming a fixed progenitor mass cutoff of $M_{\rm cut}=120~\msolar$ with explosion engines N20 and W18, are plotted as gray and dark blue hexagons.}
    \label{fig:sncc_snia_par}
\end{figure*}
The alpha elements (e.g., O, Ne, and Mg) are mainly enriched by massive stars ($\gtrsim10~\msolar$) and subsequent SNcc, as discussed by \citet{NKT13_2013_SNcc,S16,LC18}. Previous studies of NGC~3079 have compared the observed abundance pattern with an IMF-weighted SNcc yield computed with an upper progenitor mass cutoff of $\sim40$--$50~\msolar$ \citep[e.g.,][]{Kobayashi_2006,de_Plaa_2006}. Theoretical yield calculations often adopted $M_{\rm cut}=100$--$120~\msolar$ \citep[e.g.,][]{NKT13_2013_SNcc,S16,LC18}. Observationally, however, there is increasing evidence that stars with $\gtrsim18$--$30~\msolar$ may fail to produce a visible SNcc and instead collapse directly to black holes \citep{Smartt2009,Smartt_2015PASA_SNccPrognMass, JMao_2021, Fukushima_2024_M82}.

We therefore considered the IMF-weighted SNcc yields with different progenitor mass upper limits. For element $i$, the IMF-weighted SNcc yield $\langle y_i\rangle$ is
\begin{equation}
    \langle y_i\rangle_{\rm cc} = \frac{\int_{M_{\rm l}}^{M_{\rm u}} \phi(m)y_i(m)\,dm}{\int_{M_{\rm l}}^{M_{\rm u}}\phi(m)\,dm},
    \label{eq:imf_sncc_yield}
\end{equation}
where $\phi(m)$ is the IMF adopting \cite{Kroupa_2001MNRAS_IMF}. $M_{\rm l}$ and $M_{\rm u}$ are the lower and upper limits of the SNcc progenitor mass. The SNcc mass yield of element $i$ for a progenitor of mass $m$ is denoted by $y_i(m)$. We used the extrapolation and interpolation method described by \cite{Gjergo_2023ApJS..264...44G}, where the predicted metal abundances are generally consistent with Milky Way observations. We adopted the yield tables from \citet{NKT13_2013_SNcc} and \citet{LC18}. Figure~\ref{fig:sncc_snia_par} displays $M_{\rm u}=15$--$33~\msolar$ for clarity, while the quantitative scan below uses a dense grid over $M_{\rm u}=13$--$40~\msolar$. The yields from \citet{LC18} show little dependence on $M_{\rm u}$ for $M_{\rm u} \gtrsim 30~\msolar$. We further considered different initial metallicities. For \citet{NKT13_2013_SNcc}, the metallicities are given by $\rm [M/H]=\log(Z/Z_\odot) = -\infty, -1.15, -0.54, -0.24, 0.15$, and $0.55$. For the yields table from \citet{LC18}, the metallicities are $\rm [M/H] = -3, -2, -1$, and $0$. We also included the yields from \citet{S16}, considering the explosion models N20 and W18, all assuming a metallicity of $Z=0.02$ and a fixed upper progenitor mass limit of $M_{\rm u}=120~\msolar$. The abundance ratio of the SNcc theoretical prediction was then calculated via
\begin{equation}
    \frac{Z_i}{Z_j} = \left(\frac{\langle y_i\rangle_{\rm cc}/A_i}{\langle y_j\rangle_{\rm cc}/A_j}\right) \times \left(\frac{Z_{i,\odot}}{Z_{j,\odot}}\right)^{-1},
\end{equation}
where $A_i$ and $A_j$ are the atomic mass numbers for elements $i$ and $j$. $Z_{i,\odot}$ and $Z_{j,\odot}$ are the solar abundances for elements $i$ and $j$.
The results are shown in Fig.~\ref{fig:sncc_snia_par}. The O/Fe and Ne/Fe ratios from the fiducial M1 fit (dark orange star) favor the low-$M_{\rm u}$ part of the \citet{NKT13_2013_SNcc} and \citet{LC18} yield grids.

We quantified the SNcc-only constraint using the asymmetric $1\sigma$ abundance errors of M1 in Table~\ref{tab:fitpar}. Propagation of the O, Ne, and Fe errors independently gives ${\rm O/Fe}=0.47^{+0.16}_{-0.18}$ and ${\rm Ne/Fe}=0.77^{+0.33}_{-0.35}$. We evaluated the interpolated grid in $(Z,M_{\rm u})$ and defined the best-fitting $M_{\rm u}$ at the minimum error-normalized distance from these ratios. The $1\sigma$ interval was obtained by projecting the grid points whose predicted O/Fe and Ne/Fe lay within the two spectral-fitting error intervals onto the $M_{\rm u}$ axis. The \citet{NKT13_2013_SNcc} yields give $M_{\rm u}=15.3^{+2.8}_{-2.3}~\msolar$, and the \citet{LC18} yields give $M_{\rm u}=13.3^{+0.6}_{-0.3}~\msolar$.

We further considered the contribution from SNIa adopting the SNIa yields from \cite{Seitenzahl_2013_SNIa}, a 3D delayed-detonation model N100, with a progenitor metallicity of $Z=0.02$. For an SNIa fraction $f_{\rm Ia}$, the abundance ratio produced by SNIa and SNcc is
\begin{equation}
    \frac{Z_i}{Z_j} = \frac{f_{\rm Ia} y_{i,\rm Ia} + (1-f_{\rm Ia})\langle y_i\rangle_{\rm cc}}{f_{\rm Ia} y_{j,\rm Ia} + (1-f_{\rm Ia})\langle y_j\rangle_{\rm cc}} \times\frac{A_j}{A_i} \times\frac{Z_{j,\odot}}{Z_{i,\odot}},
\end{equation}
where $y_{i,\rm Ia}$ and $y_{j,\rm Ia}$ are the SNIa mass yields of elements $i$ and $j$, respectively. The three columns with nonzero SNIa contributions in Fig.~\ref{fig:sncc_snia_par} (1\%, 5\%, and 10\%) illustrate how a delayed SNIa enrichment shifts the predicted ratios.

\begin{figure}[t]
    \centering
    \includegraphics[width=0.83\linewidth]{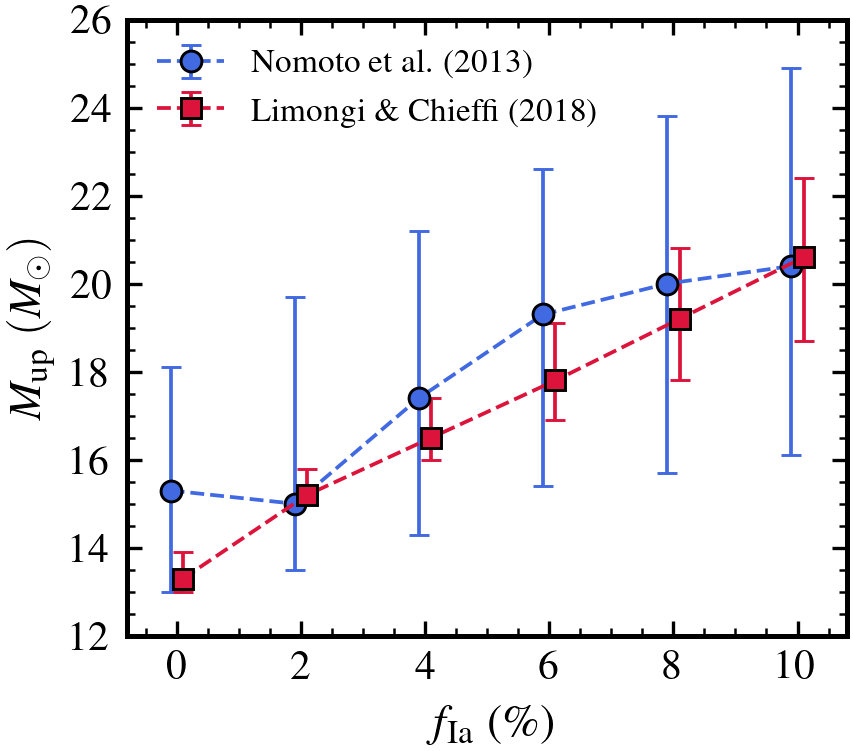}
    \caption{Best-fitting SNcc progenitor mass upper limit $M_{\rm u}$ as a function of the assumed SNIa fraction for the \citet{NKT13_2013_SNcc} and \citet{LC18} yield tables. The error bars show the $1\sigma$ ranges obtained from the M1 spectrum-fitting abundance uncertainties.}
    \label{fig:mup_snia}
\end{figure}

We applied the same grid procedure at each fixed $f_{\rm Ia}$ up to $10\%$ (Fig.~\ref{fig:mup_snia} and Table~\ref{tab:mup_snia}). The allowed $M_{\rm u}$ generally increases with the assumed SNIa fraction, while the $1\sigma$ upper limits remain below $25~\msolar$. The observed pattern remains different from the \citet{S16} predictions throughout this range. An extended scan shows that $30$--$40~\msolar$ can enter the allowed range at $f_{\rm Ia}\simeq15$--$30\%$, depending on the yield table. The differences between the \citet{NKT13_2013_SNcc} and \citet{LC18} results indicate the remaining systematic uncertainty in the SNcc yields. Future high-resolution measurements of Ne/O and Mg/O, as discussed in Sect.~\ref{sec:caveats}, will provide an additional test of the inferred progenitor mass range.

\begin{table}[t]
    \caption{SNcc progenitor mass upper limit.}
    \label{tab:mup_snia}
    \centering
    \small
    \renewcommand{\arraystretch}{1.2}
    \begin{tabular*}{\linewidth}{@{\extracolsep{\fill}}ccc@{}}
    \hline
    \hline
    $f_{\rm Ia}$ (\%) & Nomoto et al. & Limongi \& Chieffi \\
    \hline
    0  & $15.3^{+2.8}_{-2.3}$ & $13.3^{+0.6}_{-0.3}$ \\
    2  & $15.0^{+4.7}_{-1.5}$ & $15.2^{+0.6}_{-0.1}$ \\
    4  & $17.4^{+3.8}_{-3.1}$ & $16.5^{+0.9}_{-0.5}$ \\
    6  & $19.3^{+3.3}_{-3.9}$ & $17.8^{+1.3}_{-0.9}$ \\
    8  & $20.0^{+3.8}_{-4.3}$ & $19.2^{+1.6}_{-1.4}$ \\
    10 & $20.4^{+4.5}_{-4.3}$ & $20.6^{+1.8}_{-1.9}$ \\
    \hline
    \end{tabular*}
    \tablefoot{Best-fitting $M_{\rm u}$ with $1\sigma$ errors for different SNIa fractions. All masses are in $\msolar$. The two yield columns use the models of \citet{NKT13_2013_SNcc} and \citet{LC18}, respectively.}
\end{table}
Delayed SNIa enrichment may have contributed to the Fe-peak elements in the nuclear region of NGC~3079. Within the adopted \citet{NKT13_2013_SNcc} and \citet{LC18} yield frameworks, the RGS best-fit abundance ratios give $1\sigma$ upper bounds on the SNcc progenitor mass upper limit below $25~\msolar$ for $f_{\rm Ia}\leq10\%$.

\subsection{Energy drivers for NSB and GSB}
\label{sec:discussion-driver}
Before using the fitted temperatures to infer the energy drivers, we assessed the physical consistency of M2. M2 is statistically comparable to M1, although its NEIJ component requires extreme physical conditions under a single-age interpretation. We denote the X-ray-emitting volume by $V_{\rm X}$. Table~\ref{tab:fitpar} gives $\mathcal{E}_{\rm NEIJ}=n_{\rm H}n_{\rm e}V_{\rm X}=2.2\times10^{64}~\rm cm^{-3}$, $U=5.58\times10^{8}~\rm s~cm^{-3}$, and $kT_2\simeq0.10~\rm keV$. We approximated the NEIJ-emitting region as a sphere of radius $r=2~\rm kpc$, with $V_{\rm sph}=4\pi r^3/3$. The X-ray-emitting volume is $V_{\rm X}=fV_{\rm sph}$, where $f$ is the volume filling factor. We assumed a fully ionized plasma with an approximately cosmic H/He number ratio, $n_{\rm He}/n_{\rm H}\simeq0.1$. This gives $n_{\rm e}\simeq n_{\rm H}+2n_{\rm He}\simeq1.2n_{\rm H}$ and a total gas-mass density $\rho\simeq m_p(n_{\rm H}+4n_{\rm He})\simeq1.4m_pn_{\rm H}$. These assumptions give
\begin{equation}
\begin{aligned}
    &n_{\rm e} = \left(\frac{1.2\mathcal{E}_{\rm NEIJ}}{fV_{\rm sph}}\right)^{1/2}
    \simeq0.16f^{-1/2}~\rm cm^{-3},\\
    &t_{\rm ion} = \frac{U}{n_{\rm e}}
    \simeq1.1\times10^2f^{1/2}~\rm yr,\\
    &M_{\rm hot} = 1.4m_p n_{\rm H}V_{\rm X}
    =\frac{1.4m_p}{1.2}n_{\rm e}fV_{\rm sph}
    \simeq1.6\times10^8f^{1/2}~\msolar.
\end{aligned}
\end{equation}
The filling factor and volume cancel out when the mass and thermal energy are divided by $t_{\rm ion}$,
\begin{equation}
\begin{aligned}
    &\dot M_{\rm hot}=\frac{1.4m_p\mathcal{E}_{\rm NEIJ}}{U}
    \simeq1.5\times10^{6}~\msolar~\rm yr^{-1},\\
    &\dot E_{\rm e}=\frac{1.8kT_2\mathcal{E}_{\rm NEIJ}}{U}
    \simeq1.1\times10^{46}~\rm erg~s^{-1}.
\end{aligned}
\end{equation}
The inclusion of ions at the same temperature increases the heating rate to $\simeq2.1\times10^{46}~\rm erg~s^{-1}$. The conservative $1\sigma$ limits remain $\dot M_{\rm hot}\sim2\times10^{5}~\msolar~\rm yr^{-1}$ and $\dot E_{\rm e}\sim10^{45}~\rm erg~s^{-1}$. The mass-processing rate exceeds the current SFR of $2.6~\msolar~\rm yr^{-1}$ \citep{Yamagishi_2010_3079_starburst} by about five orders of magnitude. For the $2\times10^{5}~\msolar$ SMBH \citep{Kondratko_2005ApJ_ngc3079_BHmass}, the Eddington luminosity is $L_{\rm Edd}\simeq2.5\times10^{43}~\rm erg~s^{-1}$ \citep[e.g.,][]{Churazov_2005MNRAS_SMBHfeedback}, and even the conservative heating rate is higher by about 40 times. The available power from the current star formation and the AGN falls far below the M2 requirement. These results suggest that M2 serves as a phenomenological description of the unresolved ionization-age structure or a degeneracy between $U$ and normalization.

We therefore focused on M1 for the physical interpretation of the nuclear plasma. Its CX component also has a plausible physical origin. A possible H~{\sc i} outflow has been reported in NGC~3079 \citep{Shafi_2015MNRAS_3079HI}, providing cool-phase gas that can interact with highly charged ions in the X-ray--emitting plasma and produce CX emission \citep{Veilleux_2021_3079_IR}. The enhanced O~{\sc vii} He$\alpha$ forbidden emission in Fig.~\ref{fig:rgs_spec_bestfit} is consistent with this interpretation.

We assumed that the outermost X-ray-emitting gas approximately traces the forward shock and that the fitted M1 temperature measures its post-shock thermal state. The strong-shock relation $v_s=\sqrt{16kT/3\mu m_p}$ with $\mu=0.59$ and the M1 temperature, $kT_0=0.47^{+1.41}_{-0.24}~\rm keV$, give $v_s=638^{+638}_{-192}~\rm km~s^{-1}$. For an energy-conserving self-similar bubble, $v_s=3r_s/(5t)$. The adoption of $r_s=1~\rm kpc$ gives a characteristic NSB age of $0.92^{+0.39}_{-0.46}~\rm Myr$.

For a wind-driven bubble in a self-similar state that is launched by either the starburst or the SMBH, the shock velocity is related to the wind luminosity $\dot E_{\rm wind}$ by \citep{Weaver_1977ApJ_bubble, Mac_Low_1988ApJ}
\begin{equation}
   v_s = \frac35 \left(\frac{\dot E_{\rm wind}}{\rho_0}\right)^{1/5}t^{-2/5},
    \label{eq:vs}
\end{equation}
where $\rho_0$ is the CGM mass density. We write $\rho_0=\mu m_p n_0$ with $\mu=0.59$, where $n_0$ is the total particle number density of a fully ionized plasma with the cosmic composition assumed above. For an SMBH of mass $M_{\rm BH}$ in the radiatively inefficient regime, the wind luminosity is $L_{\rm BH}\approx\epsilon\dot M_{\rm BH}c^2\sim0.1\dot M_{\rm BH}c^2$, where $\epsilon\sim0.1$ is the energy-conversion efficiency for $\dot M_{\rm BH}/\dot M_{\rm Edd}\lesssim0.01$ \citep{Giustini_2019_SMBHfeedback}. Writing $L_{\rm BH}$ in terms of the Eddington luminosity $L_{\rm Edd}$ gives
\begin{equation}
\begin{aligned}
    L_{\rm BH}&=0.1\left(\frac{\dot M_{\rm BH}}{\dot M_{\rm Edd}}\right)L_{\rm Edd} \\
    &=1.38\times10^{37}~{\rm erg~s^{-1}}\times\epsilon_{\rm Edd}\left(\frac{M_{\rm BH}}{\msolar}\right)\\
\end{aligned},
\end{equation}
where $\epsilon_{\rm Edd}=\dot M_{\rm BH}/\dot M_{\rm Edd}$ is the Eddington ratio. Substituting $L_{\rm BH}$ into Eq.~\ref{eq:vs}, we obtain
\begin{equation}
    \begin{aligned}
        v_s & \approx 160~{\rm km~s^{-1}} \times \\
       & \quad \epsilon_{\rm Edd}^{1/5}\left(\frac{M_{\rm BH}}{\msolar}\right)^{1/5}\left(\frac{n_0}{10^{-3}~\rm cm^{-3}}\right)^{-1/5}\left(\frac{t}{10^6~\rm yr}\right)^{-2/5}.
    \end{aligned}
    \label{eq:vs_bh}
\end{equation}
The SMBH mass in NGC~3079 is $M_{\rm BH}\approx2\times10^{5}~\msolar$ \citep{Kondratko_2005ApJ_ngc3079_BHmass}. When we define $n_{-3}=n_0/10^{-3}~\rm cm^{-3}$ and use the M1 self-similar age above, Eq.~\ref{eq:vs_bh} gives $\epsilon_{\rm Edd}\sim4.3\times10^{-3}n_{-3}$. The large M1 temperature uncertainty broadens its coefficient to $(1.5\times10^{-3}$--$3.4\times10^{-2})$. This estimate is compatible with the measured $\epsilon_{\rm Edd}=0.01$--$0.1$ \citep{Iyomoto_2001_ApJL} for plausible ambient densities, but the energetics do not uniquely identify the SMBH as the driver.

Alternatively, if the NSB is powered by the starburst wind, the wind luminosity is \citep{Sarkar_2024_FEB_review}
\begin{equation}
    \dot E_{\rm wind} = 7\times10^{41}~\alpha~{\rm erg~s^{-1}}\frac{\rm SFR}{\msolar~\rm yr^{-1}},
\end{equation}
where $\alpha$ is the energy-loading factor. From Eq.~\ref{eq:vs}, the shock velocity generated is
\begin{equation}
    \begin{aligned}
         v_s & \approx 1409~{\rm km~s^{-1}}\times \\
         \quad & \left(\frac{\alpha \mathrm{SFR}}{\msolar~\rm yr^{-1}}\right)^{1/5} \left(\frac{n_0}{10^{-3}~\rm cm^{-3}}\right)^{-1/5} \left(\frac{t}{10^6~\rm yr}\right)^{-2/5}.
    \end{aligned}
\end{equation}
For $\alpha=0.3$ \citep{Strickland_2009ApJ_SNe, Gentry_2017MNRAS.465.2471G_SNe, Vasiliev_2017MNRAS.468.2757V_SNe, Fielding_2018MNRAS.481.3325F_SNe}, an $\mathrm{SFR}\gtrsim0.054~n_{-3}~\msolar~\rm yr^{-1}$ is required to produce the M1 NSB shock. The M1 temperature uncertainty allows a broad coefficient range of $0.018$--$0.43$. These values are below the infrared-derived $\mathrm{SFR}=2.6~\msolar~\rm yr^{-1}$ \citep{Yamagishi_2010_3079_starburst} over a plausible range of $n_0$, so the nuclear starburst can also supply the required mechanical power.

The M1 energy scale is insufficient to distinguish the drivers of the NSB, and both AGN and starburst events may play a role. In addition, the HCO$^{+}$/HCN ratio derived from millimeter observations exhibits a composite signature characteristic of both AGN and starburst activity \citep{Li_2019MNRAS_3079_mm_obs}. While \cite{Li_2024ApJ} estimated an engine power of $\gtrsim10^{42}~\rm erg~s^{-1}$, our results are compatible with this estimate because their analysis assumed a unity volume-filling factor and therefore effectively provided an upper limit.

Similarly, the shock velocity of the GSB can be derived from the best-fit temperature in the combined NE+SW sectors (Table~\ref{tab:fitpar}), yielding $v_s=506^{+55}_{-25}~\rm km~s^{-1}$. For the same energy-conserving self-similar solution as used for the NSB, the dynamical age is $t=3r_s/(5v_s)$. A radius of $r_{\rm s}=5'\approx25~\rm kpc$ gives $t=29.6^{+1.6}_{-2.9}~\rm Myr$, comparable to the $\sim10^7~\rm yr$ sputtering timescale for dust grains entrained in the outflow \citep{Veilleux_2021_3079_IR}. The errors of the characteristic timescale may be underestimated given the current assumptions. At fixed radius and ambient density, the age and required mechanical power scale with the NE+SW temperature through $t\propto(kT)^{-1/2}$ and $\dot E_{\rm wind}\propto v_s^{3}\propto(kT)^{3/2}$. The temperature is less sensitive than the abundance ratios in the model tests considered here. Addition of CX in the nucleus (M3 to M1) leaves the thermal temperature at $0.46$--$0.47~\rm keV$ while O/Fe changes by a factor of $3.4$, and the specific simulations in Appendix~\ref{sec:bias} bias single-temperature fits by $\sim20\%$ in temperature.
To produce the observed shock, a star formation rate of $\mathrm{SFR}\gtrsim17~n_{-3}~\msolar~\rm yr^{-1}$ would be required. For the fiducial $n_{-3}=1$, this value exceeds the current star formation activity in NGC~3079, suggesting that the GSB is unlikely to be associated with the ongoing nuclear starburst and instead could have originated from another feedback event about $30~\rm Myr$ ago. This might indicate an episodic feedback cycle in NGC~3079, as seen in numerical simulations of disk galaxies \citep{Pillepich_2021_TNG}.

\subsection{Caveats and future perspective}
\label{sec:caveats}
\begin{figure*}
    \centering
    \includegraphics[width=0.95\linewidth,trim={2cm 0cm 2cm 0.75cm},clip]{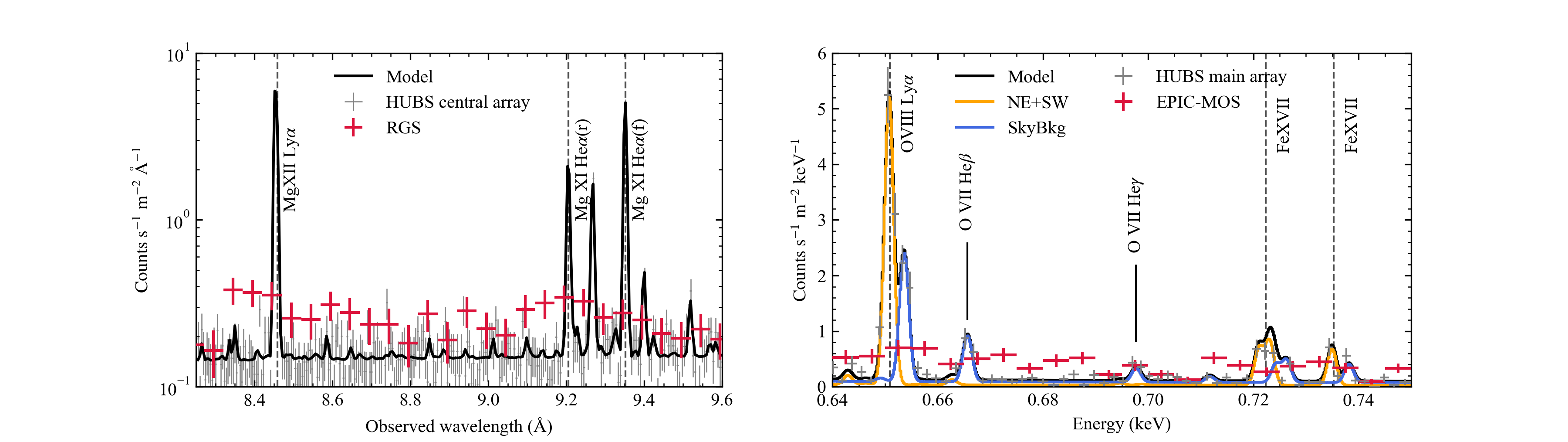}
    \caption{\textit{Left}: Mock 200 ks HUBS central array spectrum for the nuclear region in NGC 3079. The Mg {\sc xii} Ly$\alpha$, Mg {\sc xi} He$\alpha$-z, and Mg {\sc xi} He$\alpha$-w emission lines are marked and labeled. The gray and red crosses and black curves show the mock and observed data and spectral model M1, respectively. \textit{Right}: Mock 200 ks HUBS main-array spectrum for the NE and SW region with prominent lines labeled. The gray and red crosses show the mocked HUBS and observed EPIC-MOS spectrum. The black, orange, and blue curves show the total model, the combined NE+SW emission, and the soft X-ray sky background, respectively.}
    \label{fig:mock_hubs}
\end{figure*}
Ideally, the Ne/O and Mg/O ratios serve as more reliable tracers of SNcc enrichment because O, Ne, and Mg are primarily synthesized in SNcc \citep{Simionescu_2019MNRAS, Mernier_chem_enrich_inIGrCM}. In contrast, Fe is mainly produced by SNIa \citep[e.g.,][]{Iwamoto_1999_SNIa, Seitenzahl_2013_SNIa}, making the interpretation of $\alpha$/Fe ratios more complex. However, the effective area of the RGS instrument drops significantly below $10~\ang$, resulting in large uncertainties in the Mg abundance (see Table~\ref{tab:fitpar}). Future missions with non-dispersive high-resolution X-ray spectrometers, such as the Hot Universe Baryon Surveyor (HUBS; \citealt{cui_HUBS}) and the New Advanced Telescope for High-ENergy Astrophysics (NewAthena; \citealt{NewAthena}), are expected to provide more robust constraints on the Ne, Mg, and Si abundances.

The nuclear region of NGC 3079 has a spatial extension of $\sim1'$ in diameter and is suitable for HUBS observations using the central array with a field of view of $3'\times3'$. A comparison of the simulated 200 ks HUBS central array spectrum of the nuclear region and the observed RGS spectrum is shown in the left panel of Fig.~\ref{fig:mock_hubs}. The Mg {\sc xii} Ly$\alpha$, Mg {\sc xi} He$\alpha$ resonance and forbidden lines are prominent in the mock spectrum. This enables us to measure the abundance more precisely and allows us to directly compare the Ne/O and Mg/O ratios with SNcc yield predictions, almost independent of SNIa contamination. The right panel of Fig.~\ref{fig:mock_hubs} shows the mock 200 ks HUBS main-array spectrum for the NE and SW sectors, where the O \textsc{viii} Ly$\alpha$ line and the Fe \textsc{xvii} complex are clearly resolved. The enhanced spectral quality with high-resolution microcalorimeters enables us to use more sophisticated spectral models incorporating temperature and abundance structures, thereby providing tighter constraints on the elemental abundance ratios.

Analyses of RGS data have reported supersolar O/Fe in M82 and M51, which is higher than the value obtained with our best-fit model for NGC~3079 \citep{Zhang_2014_M82,Zhang_2022_M51}. The M82 analyses adopted either a single-temperature CIE+CX model \citep{Zhang_2014_M82} or a multi-temperature thermal model including a cool CIE component without CX \citep{Fukushima_2024_M82}. Differences in the temperature structure and CX treatment can affect the inferred abundance ratios, as illustrated by our model comparisons in Table~\ref{tab:fitpar}. However, different chemical enrichment histories might also contribute to the contrast between galaxies. Delayed SNIa enrichment provides one possible explanation, as explored in Sect.~\ref{sec:abund:prog}. The X-ray-emitting gas might also contain preexisting disk material that is heated or entrained by the outflow. Its abundance pattern and the extent to which newly synthesized metals mix into the observed hot phase can affect the measured ratios. A consistently analyzed sample of galaxies, together with constraints on their star formation histories and preexisting chemical composition, is needed to distinguish variations in enrichment and mixing from differences in the spectral modeling.

\section{Conclusion}
We presented \textit{XMM-Newton} RGS (nuclear region, NSB included) and EPIC-MOS (GSB) spectroscopic constraints on the hot gas in NGC~3079. Our results are listed below.
\begin{itemize}
    \item The Compton-thick AGN contributes negligibly in the RGS band, and prominent emission lines are observed in the nuclear region.
    \item The fiducial and best-supported nuclear region model is mCIE+CX (M1), in which CX contributes $\approx13\%$ to the observed $0.2$--$2~\rm keV$ energy flux. The best-fit mCIE parameters are $kT_0=0.47^{+1.41}_{-0.24}~\rm keV$ and $\sigma_T=0.76^{+0.24}_{-0.27}$. Although NEIJ+mCIE (M2) is statistically comparable, its fitted emission measure and ionization age require physically implausible mass-processing and heating rates under a single-age interpretation.
    \item The fitted O/Fe and Ne/Fe ratios are $0.47$ and $0.77$ under M1, in which CX contributes strongly to the O features, but little to the Fe~{\sc xvii} bands.
    \item The comparison of the M1 ratios with IMF-weighted SNcc yields gives $M_{\rm u}\simeq13$--$15~\msolar$ for pure SNcc enrichment and $1\sigma$ upper limits below $25~\msolar$ for SNIa fractions up to $10\%$.
    \item The GSB (out to $r\sim5'\approx25~\rm kpc$) is adequately described by a single-temperature \texttt{CIE} model given the available photon statistics, with a nearly constant temperature of $0.30$--$0.38~\rm keV$.
    \item The M1 temperature gives $t_{\rm NSB}=0.92^{+0.39}_{-0.46}~\rm Myr$, whereas $t_{\rm GSB}\sim30~\rm Myr$. While either the AGN or the current nuclear starburst can power the NSB, reproducing the GSB shock would require $\mathrm{SFR}\gtrsim17~n_{-3}~\msolar~\rm yr^{-1}$, which is above the present $\mathrm{SFR}\approx2.6~\msolar~\rm yr^{-1}$ for the fiducial $n_{-3}=1$, favoring previous feedback events. The coexistence of superbubbles on different scales in NGC~3079 might indicate that an episodic feedback process operates on a timescale of $\sim30~\rm Myr$.
\end{itemize}

\section*{Data availability}
The processed spectra and response files, SPEX fitting recipes, imaging products, plotting data, and reproduction scripts are available at Zenodo\footnote{\url{https://doi.org/10.5281/zenodo.23280068}}.

\begin{acknowledgements}
We thank the anonymous referee for helpful comments and suggestions. This work was supported in part by the National Natural Science Foundation of China through Grants 11821303, 12273112, and E3GJ251110, and by the Ministry of Science and Technology of China through Grant 2018YFA0404502. J.~M. acknowledges support from the Tsinghua Dushi Program 53121200126. We acknowledge the Tsinghua Astrophysics High-Performance Computing (TAHPC) platform for providing computational and data storage resources.
\end{acknowledgements}

\bibliography{ref}{}
\bibliographystyle{aa}

\begin{appendix}
\nolinenumbers

\twocolumn[{
\section{Spectral modeling of the X-ray foreground and background}
\label{appendix:xray_background}
\begin{minipage}{\textwidth}
    \centering
    \includegraphics[width=\linewidth, trim={1.5cm 0cm 1.5cm 1cm}]{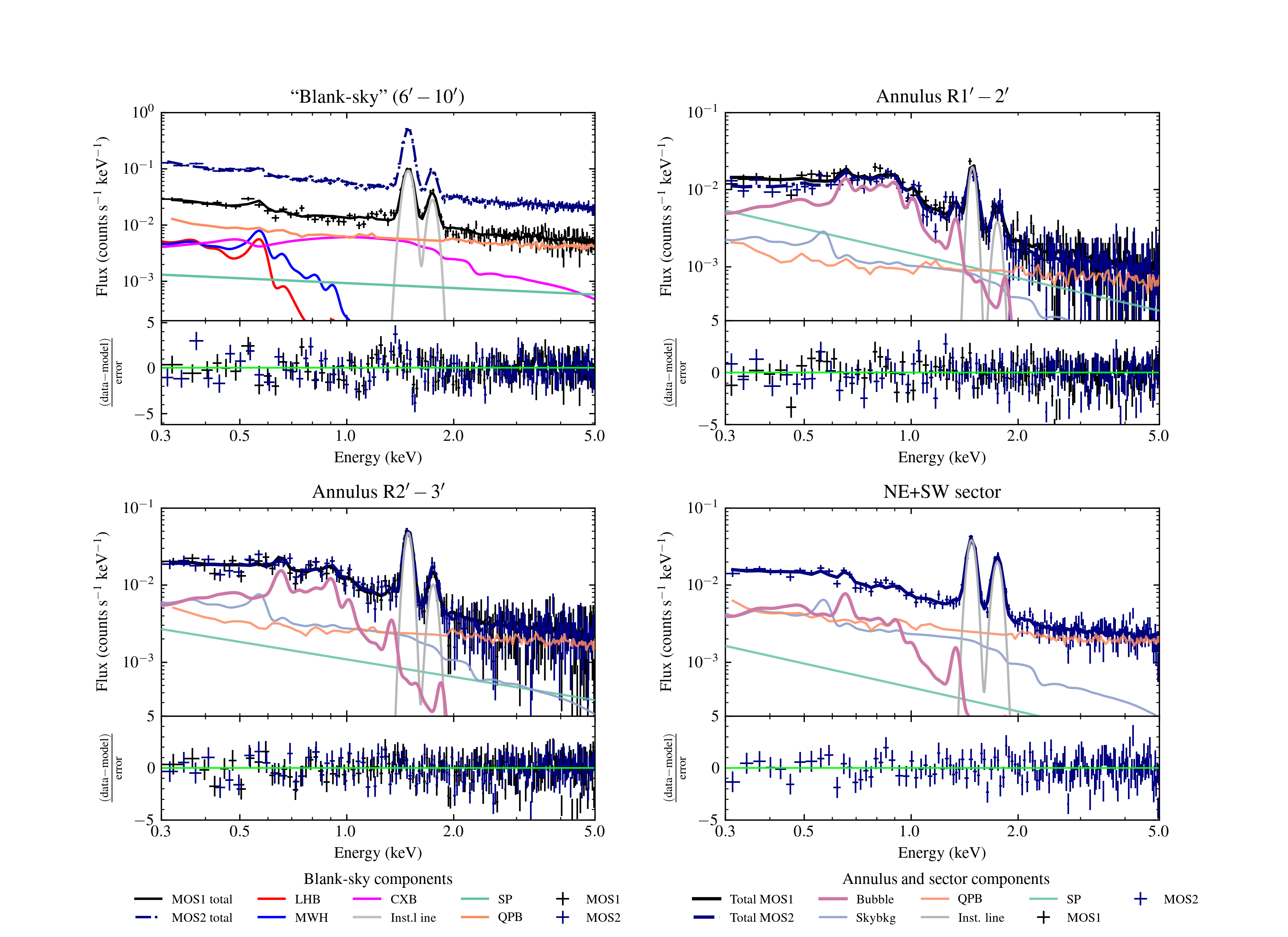}
    \captionof{figure}{Fitted \textit{XMM-Newton} MOS spectra. Each panel shows the MOS1 (black crosses) and MOS2 (navy crosses) data, the best-fit total models (MOS1: solid black; MOS2: navy dashed-dotted), and the residuals in the lower sub-panels.
    \textit{Upper left}: Fit to the combined $6'\!-\!10'$ ``blank-sky'' region. The additive background components are LHB (red), MWH (blue), CXB (magenta), residual soft proton SP (green), QPB (orange), and instrumental fluorescence lines (Inst; gray). For clarity, only one set of component curves is shown, while both MOS1 and MOS2 total models are plotted.
    \textit{Upper right, lower left, lower right}: Fits to the $1'\!-\!2'$ annulus, $2'\!-\!3'$ annulus, and the NE--SW sector, respectively. In these source regions, the re-normalized sky-background component (Skybkg; light blue) is included together with the hot-gas CIE component representing the GSB ``Bubble'' emission (violet), QPB (orange), SP (green), and instrumental lines (Inst.\ line; gray).}
    \label{fig:fit_epic}
\end{minipage}
\vspace{1em}
}]
We extracted a MOS spectrum from the $6'\text{--}10'$ annulus as a blank-sky region to constrain the soft X-ray foreground and background components, including the Local Hot Bubble (LHB), Milky Way halo (MWH), and cosmic X-ray background (CXB). We modeled these with an unabsorbed CIE (LHB), an absorbed CIE (MWH), and an absorbed power law (CXB), fixing the neutral hydrogen column density to $N_{\rm H}=8.78\times10^{19}~\rm cm^{-2}$ and adding two Gaussians for the instrumental Al~K$\alpha$ and Si~K$\alpha$ lines plus an additional power law for residual SP.
Because the NGC~3079 observations were taken near solar minimum and show no obvious solar-wind charge-exchange features, we did not include such a component.

The best-fit LHB temperature is $0.11\pm0.02~\rm keV$, consistent with $0.097\pm0.013$ from \citet{Liu_2017ApJ_LHB}. The MWH temperature is $0.15^{+0.08}_{-0.00}~\rm keV$ (pegged at the lower bound), with abundance fixed to $Z=0.3~Z_\odot$ \citep{Nicastro_2023ApJ}. The CXB photon index is $\Gamma_{\rm CXB}=1.49\pm0.08$, consistent with $1.41\pm0.06$ from \citet{DeLuca_2004A&A...419..837D}. The fit is shown in the upper left panel of Fig.~\ref{fig:fit_epic}.

\FloatBarrier

\section{Temperature and abundance bias of the single-temperature model}
\label{sec:bias}
A single-temperature fit to a multi-temperature plasma biases the fitted abundances. Similar effects have been extensively reported in studies of the intracluster and intragroup medium known as the ``Fe bias'' and ``inverse Fe bias'' \citep{Gastaldello_2010A&A_InverseFebias}. These biases arise from the dominant statistical weight of the Fe-L complex in spectral fitting, leading to systematic under- or overestimation of Fe abundances in multi-temperature plasmas across typical cluster and group temperature ranges. In galactic environments, where plasma temperatures are lower, the fitting of temperature and abundances becomes dependent on both the Fe L-shell complex and prominent emission lines such as O {\sc vii} and O {\sc viii}. When using a uniform temperature model, the temperature is artificially increased to match the bremsstrahlung tail between $1$--$2~\rm keV$, produced by the hotter plasma phase. This leads to a lower Fe abundance in an attempt to suppress the enhanced Fe L-shell emission around $E\sim 0.7$--$1.1~\rm keV$ due to temperature elevation. This effect can potentially bias the O/Fe and Ne/Fe ratios.

\begin{figure}[!ht]
    \centering
    \includegraphics[width=\linewidth, trim={3.5cm 1cm 2cm 1cm}, clip]{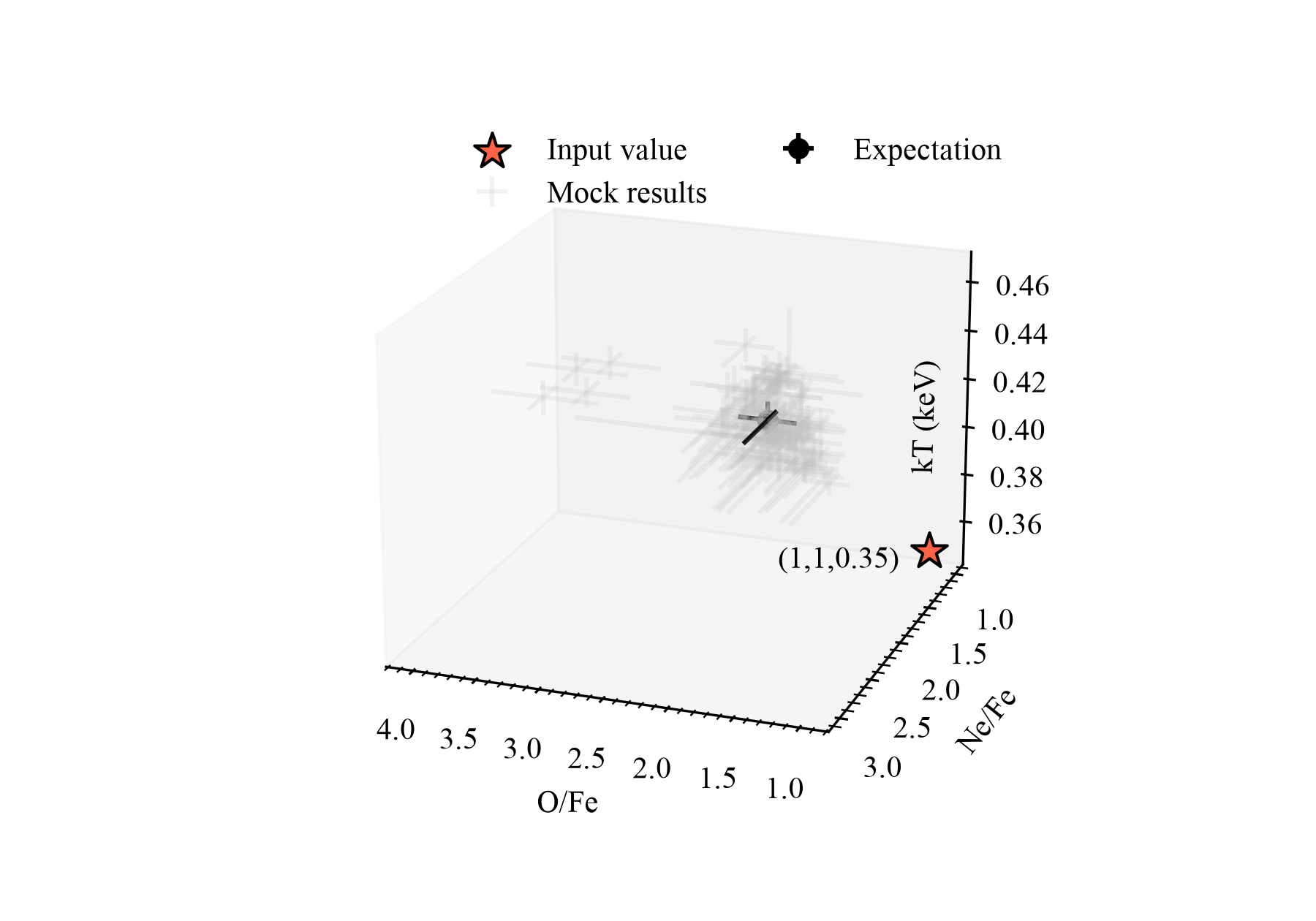}
    \caption{Fitted temperature vs O/Fe vs Ne/Fe ratios to the mocked multi-temperature spectrum with a single temperature model. The dark-orange star shows the input value $({\rm O/Fe,Ne/Fe},kT_0)$ = $(1,1,0.35~\rm keV)$ of the multi-temperature model, where the central temperature $kT_0$ is plotted. The silver data points show the fitted values of each mocked spectrum. The black cross shows the expectation of all the mock spectra with statistical errors.}
    \label{fig:fit_mock}
\end{figure}

To illustrate this bias, we simulated 100 MOS2 spectra using SPEX for a multi-temperature CIE model, where the central temperature $kT_0=0.35~\rm keV$, variance $\sigma=0.4$, and $Z=1~Z_\odot$, and fit them with a single-temperature \texttt{CIE}. As shown in Fig.~\ref{fig:fit_mock}, the fitted O/Fe and Ne/Fe ratios are systematically higher than the input values. The fitted temperature has an expectation and statistical errors of $kT_0=0.420^{+0.008}_{-0.006}~\rm keV$ and is elevated.

\FloatBarrier

\section{Influence of CX on metal abundances}
\label{sec:discuss_cx}
\begin{figure}[!ht]
    \centering
    \includegraphics[width=\linewidth]{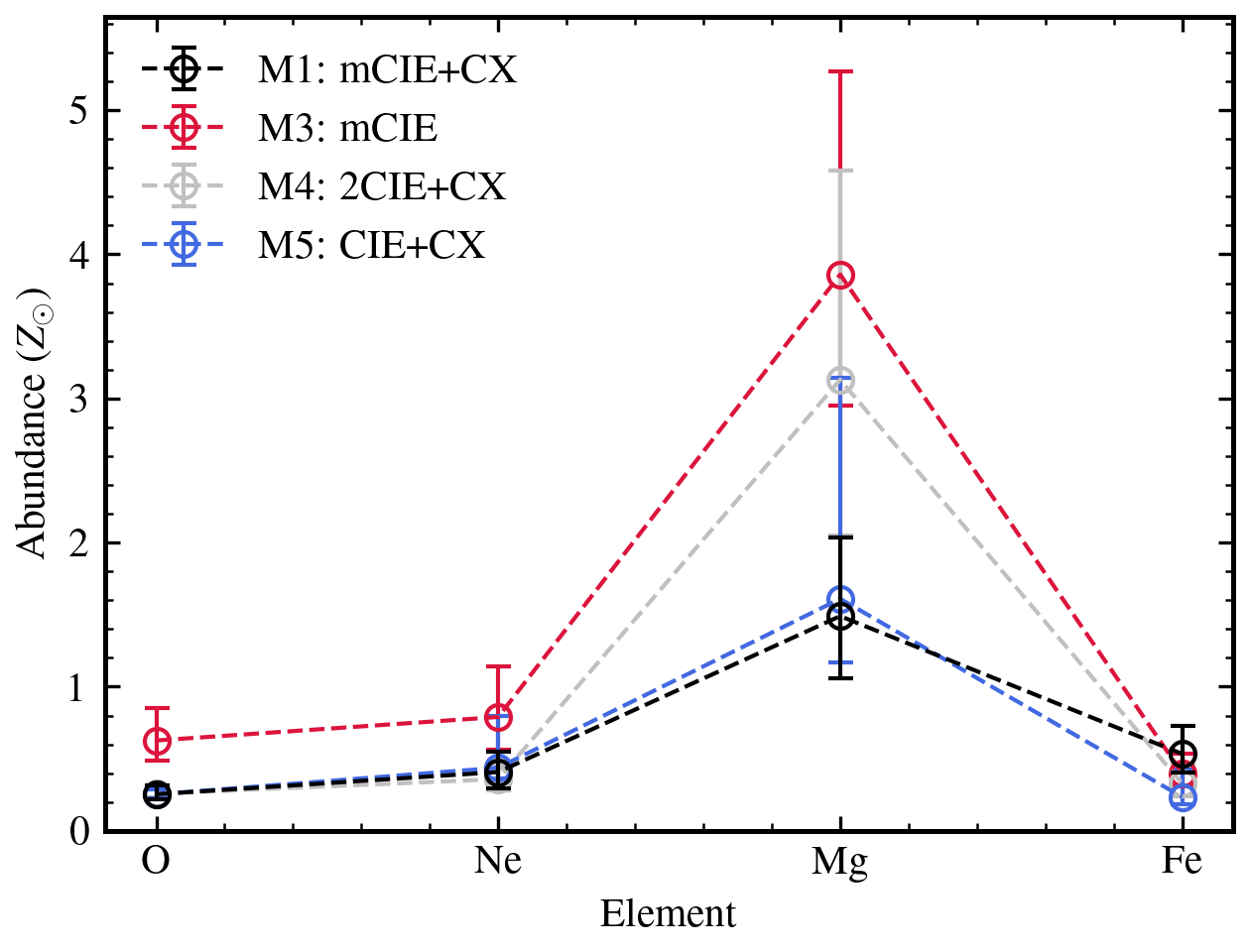}
    \caption{Comparison of the elemental abundances obtained with the CX-related models listed in Table~\ref{tab:fitpar}. M1 (mCIE+CX), M3 (mCIE without CX), M4 (2CIE+CX), and M5 (CIE+CX) are shown as black, red, silver, and blue circles, respectively.}
    \label{fig:compare_cx}
\end{figure}
CX contamination complicates the analysis of the target spectrum \citep{Zhang_2014_M82, Gu_2016A&A_cx}. As shown in Table~\ref{tab:fitpar}, adding a CX component increases the best-fit temperature of the \texttt{CIE} plasma, and Fig.~\ref{fig:compare_cx} shows that it leads to lower inferred $\alpha$-element abundances. In our case, the prominent O~{\sc vii} He$\alpha$ forbidden emission motivates the inclusion of CX (Fig.~\ref{fig:rgs_spec_bestfit}). The higher-order O~{\sc viii} Lyman lines are blended with Fe-L emission and are not used as independent evidence for CX. Without a CX component, a lower plasma temperature is required to reproduce the observed O~{\sc vii}/O~{\sc viii} ratio. Because CX adds flux to lines such as O~{\sc viii} Ly$\alpha$, Ne~{\sc x} Ly$\alpha$, and Mg~{\sc xi} He$\alpha$, the inferred $\alpha$-element abundances are correspondingly reduced. In contrast, the Fe abundance is largely unaffected because the Fe-L emission is dominated by the \texttt{CIE} component.

\begin{figure*}
    \centering
    \includegraphics[width=\linewidth]{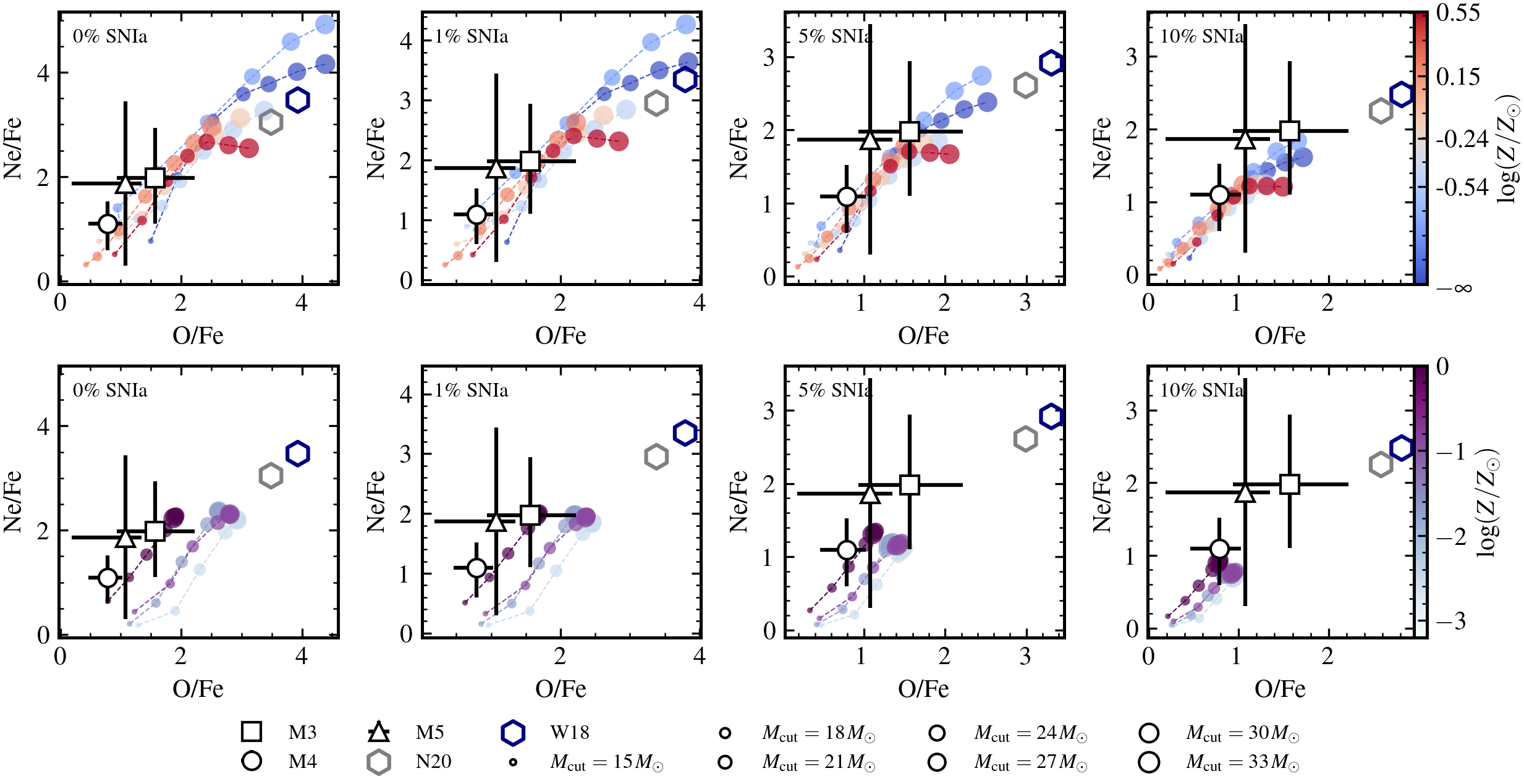}
    \caption{Same as Fig.~\ref{fig:sncc_snia_par} but comparing the fitted results of M3 (open square), M4 (open circle), and M5 (open triangle) in Table~\ref{tab:fitpar} to the theoretical predictions.}
    \label{fig:sncc_snia_par_m345}
\end{figure*}

M2 is statistically comparable to M1 ($\Delta{\rm AIC_c}=0.8$), whereas M3--M5 are considerably less supported by the data. As discussed in Sect.~\ref{sec:discussion-driver}, the best-fit M2 NEIJ component is not physically viable as a single-age plasma, so the main SNcc inference uses M1 alone. For completeness, Fig.~\ref{fig:sncc_snia_par_m345} shows how the abundance ratios from the alternative CIE and CX prescriptions M3--M5 would alter the progenitor mass comparison if treated only as spectral-model systematics. No evidence is found for an upper mass cutoff of SNcc progenitors exceeding $M_{\rm u} \approx 27$--$30~\msolar$ when compared with pure SNcc yield predictions from \citet{NKT13_2013_SNcc} and \citet{LC18}. The non-zero SNIa fractions illustrated here extend only to $10\%$; larger fractions can admit higher cutoffs, as discussed for M1 in Sect.~\ref{sec:abund:prog}. At larger SNIa fractions, some of the fitted ratios lie beyond the \citet{LC18} model grid, while the \citet{LC18} yields become nearly indistinguishable for $M_{\rm u} \gtrsim 30~\msolar$.

\FloatBarrier

\section{Other SNIa models}
\label{sec:append:snia_model}
\begin{figure}[!ht]
    \centering
    \includegraphics[width=\linewidth,trim={5mm 2mm 4mm 6.5mm},clip]{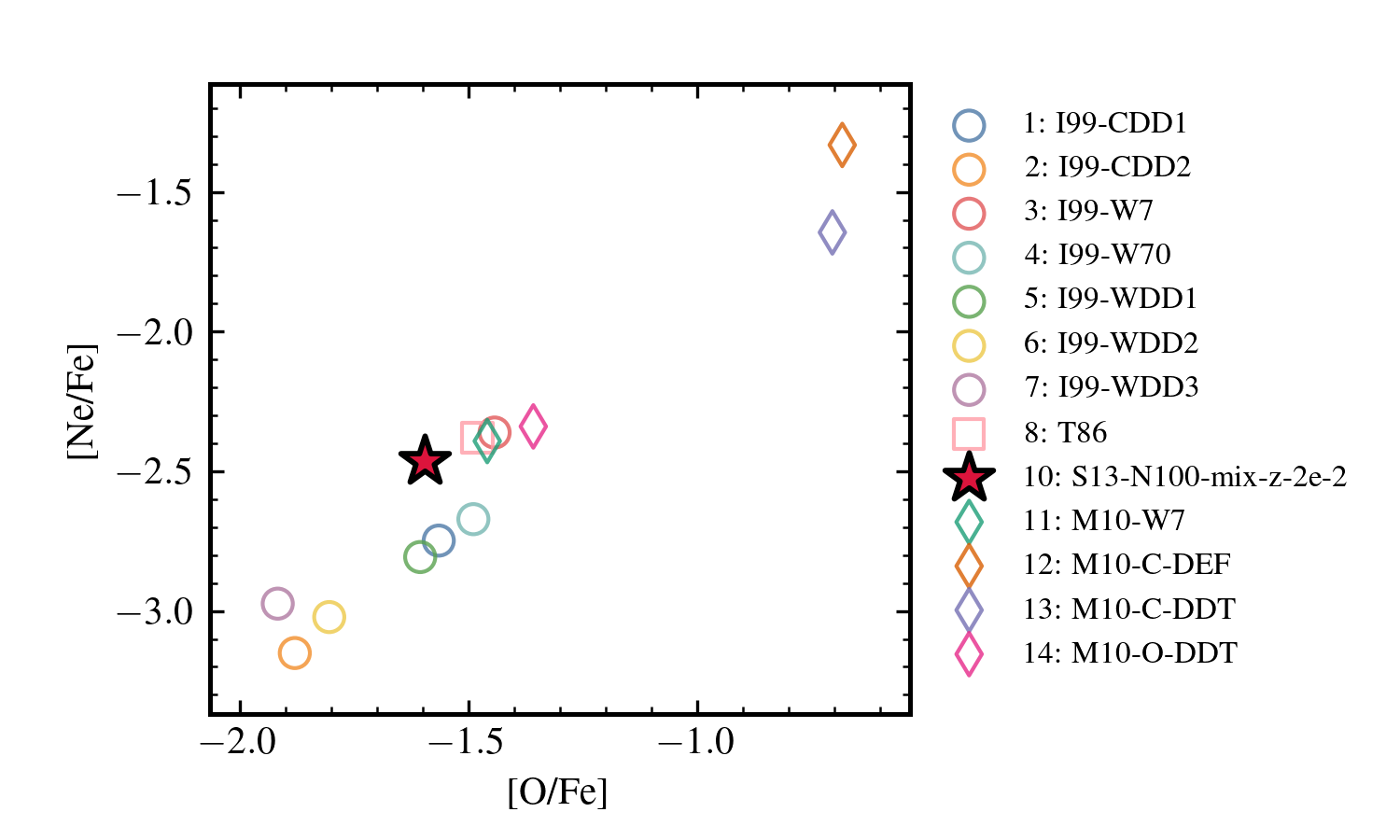}
    \caption{[O/Fe] and [Ne/Fe] ratios of different SNIa theoretical models. We highlight the SNIa model used in the main text, Fig.~\ref{fig:sncc_snia_par}, and Fig.~\ref{fig:sncc_snia_par_m345} as the red star. Details of the models are described in the text of Appendix~\ref{sec:append:snia_model}.}
    \label{fig:snia_different_models}
\end{figure}

\begin{figure*}
    \centering
    \includegraphics[width=\linewidth]{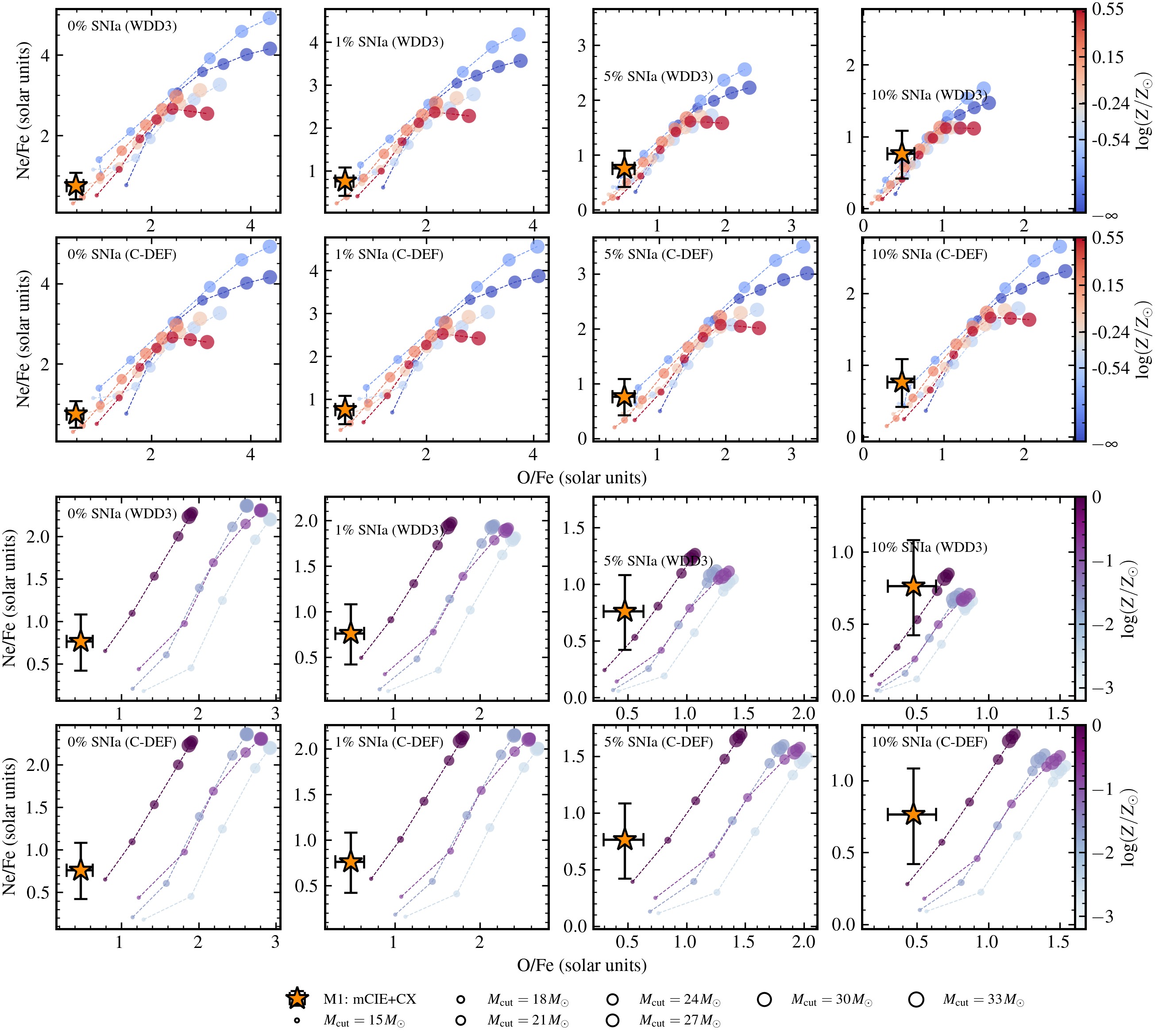}
    \caption{Comparison of the observed O/Fe and Ne/Fe ratios with SNcc yields combined with alternative SNIa yields. The top and bottom two rows adopt SNcc tables from \citet{NKT13_2013_SNcc} and \citet{LC18}, respectively. The first and third rows adopt the \cite{Iwamoto_1999_SNIa} WDD3 SNIa yields, while the second and fourth rows adopt the \cite{Maeda_2010} C-DEF yields. The orange star shows the fiducial M1 measurement; other labels and colors are identical to Fig.~\ref{fig:sncc_snia_par}.}
    \label{fig:sncc_snia_othermodel}
\end{figure*}

Figure~\ref{fig:snia_different_models} shows the [O/Fe] and [Ne/Fe] ratios predicted by different theoretical SNIa models, where [X/Y] is defined as
\begin{equation}
    \mathrm{[X/Y]} = \log_{10}\left(\frac{n_{\rm X,i}/n_{\rm X,\odot}}{n_{\rm Y,i}/n_{\rm Y,\odot}}\right)
,\end{equation}
where $n_{\rm X,i}$ and $n_{\rm Y,i}$ are the number abundances of elements X and Y in the target, and $n_{\rm X,\odot}$ and $n_{\rm Y,\odot}$ are those in the Sun. The SNIa models shown in Fig.~\ref{fig:snia_different_models} include \citet[][I99]{Iwamoto_1999_SNIa}: the pure-deflagration models W7 and W70, and the delayed-detonation models CDD1--2 and WDD1--3; \citet[][T86]{T86}: the fast-deflagration model W7; \cite{T93}: the deflagration model W7; \citet[][S13]{Seitenzahl_2013_SNIa}: the 3D delayed-detonation model N100 with a progenitor metallicity of 0.02, for which the yields are insensitive to metallicity over the range considered; and \citet[][M10]{Maeda_2010}: the 2D delayed-detonation models C-DEF, C-DDT, and O-DDT.

In Fig.~\ref{fig:sncc_snia_othermodel}, we compare the observed O/Fe and Ne/Fe ratios with theoretical predictions obtained using the SNcc models and two alternative SNIa models: WDD3 from \cite{Iwamoto_1999_SNIa} (index 7) and C-DEF from \cite{Maeda_2010} (index 12). These two models give the lowest and highest O/Fe ratios, respectively. In each panel, the extrapolation and interpolation method described by \citet{Gjergo_2023ApJS..264...44G} is used. A higher O/Fe (and Ne/Fe) ratio in the SNIa yields shifts the combined prediction toward the upper left, leading to a lower inferred upper-mass cutoff for SNcc progenitors.

\end{appendix}

\end{document}